\input lanlmac

\def\quad{{\ \ }}

\def\ads{$AdS_5\times S^5$}
\def\op{${\cal O}_\Sigma$}

\def\Tr{{\rm Tr}}

\def\BMN{{\rm BMN}}

\def\ope{{\cal O}_{\Sigma}}

\def\cD{{\cal D}}
\def\cL{{\cal L}}
\def\cN{{\cal N}}
\def\cO{{\cal O}}
\def\cS{{\cal S}}
\def\vev#1{{\left<#1\right>}}

\let\includefigures=\iftrue
\newfam\black
\includefigures
\input epsf
\def\figin{\epsfcheck\figin}\def\figins{\epsfcheck\figins}
\def\epsfcheck{\ifx\epsfbox\UnDeFiNeD
\message{(NO epsf.tex, FIGURES WILL BE IGNORED)}
\gdef\figin##1{\vskip2in}\gdef\figins##1{\hskip.5in}
\else\message{(FIGURES WILL BE INCLUDED)}%
\gdef\figin##1{##1}\gdef\figins##1{##1}\fi}
\def\DefWarn#1{}
\def\figinsert{\goodbreak\midinsert}
\def\ifig#1#2#3{\DefWarn#1\xdef#1{fig.~\the\figno}
\writedef{#1\leftbracket fig.\noexpand~\the\figno}%
\figinsert\figin{\centerline{#3}}\medskip\centerline{\vbox{\baselineskip12pt
\advance\hsize by -1truein\noindent\footnotefont{\bf Fig.~\the\figno:}
#2}}
\bigskip\endinsert\global\advance\figno by1}
\else
\def\ifig#1#2#3{\xdef#1{fig.~\the\figno}
\writedef{#1\leftbracket fig.\noexpand~\the\figno}%
#2}}
\global\advance\figno by1}
\fi


\def\sym{ \> {\vcenter {\vbox
{\hrule height.6pt
\hbox {\vrule width.6pt height5pt
\kern5pt
\vrule width.6pt height5pt
\kern5pt
\vrule width.6pt height5pt}
\hrule height.6pt}
}
} \>
}
\def\fund{ \> {\vcenter {\vbox
{\hrule height.6pt
\hbox {\vrule width.6pt height5pt
\kern5pt
\vrule width.6pt height5pt }
\hrule height.6pt}
}
} \>
}
\def\anti{ \> {\vcenter {\vbox
{\hrule height.6pt
\hbox {\vrule width.6pt height5pt
\kern5pt
\vrule width.6pt height5pt }
\hrule height.6pt
\hbox {\vrule width.6pt height5pt
\kern5pt
\vrule width.6pt height5pt }
\hrule height.6pt}
}
} \>
}


\lref\GomisFI{
J.~Gomis and S.~Matsuura,
``Bubbling surface operators and S-duality,''
JHEP {\bf 0706}, 025 (2007)
[arXiv:0704.1657 [hep-th]].
}

\lref\MaldacenaRE{
J.~M.~Maldacena,
``The large $N$ limit of superconformal field theories and supergravity,''
Adv.\ Theor.\ Math.\ Phys.\ {\bf 2}, 231 (1998)
[Int.\ J.\ Theor.\ Phys.\ {\bf 38}, 1113 (1999)]
[arXiv:hep-th/9711200].
}

\lref\WittenQJ{
E.~Witten,
``Anti-de Sitter space and holography,''
Adv.\ Theor.\ Math.\ Phys.\ {\bf 2}, 253 (1998)
[arXiv:hep-th/9802150].
}

\lref\GubserBC{
S.~S.~Gubser, I.~R.~Klebanov and A.~M.~Polyakov,
``Gauge theory correlators from non-critical string theory,''
Phys.\ Lett.\ B {\bf 428}, 105 (1998)
[arXiv:hep-th/9802109].
}

\lref\LinNH{
H.~Lin and J.~M.~Maldacena,
``Fivebranes from gauge theory,''
Phys.\ Rev.\ D {\bf 74}, 084014 (2006)
[arXiv:hep-th/0509235].
}

\lref\BianchiKW{
M.~Bianchi, D.~Z.~Freedman and K.~Skenderis,
``Holographic renormalization,''
Nucl.\ Phys.\ B {\bf 631}, 159 (2002)
[arXiv:hep-th/0112119].
}

\lref\LinNB{
H.~Lin, O.~Lunin and J.~M.~Maldacena,
``Bubbling $AdS$ space and $1/2$ BPS geometries,''
JHEP {\bf 0410}, 025 (2004)
[arXiv:hep-th/0409174].
}

\lref\SkenderisUY{
K.~Skenderis and M.~Taylor,
``Kaluza-Klein holography,''
JHEP {\bf 0605}, 057 (2006)
[arXiv:hep-th/0603016].
}

\lref\SkenderisYB{
K.~Skenderis and M.~Taylor,
``Anatomy of bubbling solutions,''
JHEP {\bf 0709}, 019 (2007)
[arXiv:0706.0216 [hep-th]].
}

\lref\Nieuwen{
H.~J.~Kim, L.~J.~Romans and P.~van Nieuwenhuizen,
``The mass spectrum of chiral $\cN=2$ $D=10$ supergravity on $S^5$,''
Phys.\ Rev.\ D {\bf 32}, 389 (1985).
}

\lref\GukovJK{
S.~Gukov and E.~Witten,
``Gauge theory, ramification, and the geometric langlands program,''
arXiv:hep-th/0612073.
}

\lref\KapustinPY{
A.~Kapustin,
``Wilson-'t~Hooft operators in four-dimensional gauge theories and
S-duality,''
Phys.\ Rev.\ D {\bf 74}, 025005 (2006)
[arXiv:hep-th/0501015].
}

\lref\KapustinPK{
A.~Kapustin and E.~Witten,
``Electric-magnetic duality and the geometric Langlands program,''
arXiv:hep-th/0604151.
}

\lref\BMN{
D.~E.~Berenstein, J.~M.~Maldacena and H.~S.~Nastase,
``Strings in flat space and pp waves from $\cN = 4$ super Yang Mills,''
JHEP {\bf 0204}, 013 (2002)
[arXiv:hep-th/0202021].
}

\lref\GrahamW{
C.~R.~Graham and E.~Witten,
``Conformal anomaly of submanifold observables in $AdS$/CFT correspondence,''
Nucl.\ Phys.\ B {\bf 546}, 52 (1999)
[arXiv:hep-th/9901021].
}

\lref\Constable{
N.~R.~Constable, J.~Erdmenger, Z.~Guralnik and I.~Kirsch,
``Intersecting D3-branes and holography,''
Phys.\ Rev.\ D {\bf 68}, 106007 (2003)
[arXiv:hep-th/0211222].
}

\lref\ZaremboWL{
K.~Zarembo,
``Supersymmetric Wilson loops,''
Nucl.\ Phys.\ B {\bf 643}, 157 (2002)
[arXiv:hep-th/0205160].
}

\lref\SchwimmerYH{
A.~Schwimmer and S.~Theisen,
``Entanglement entropy, trace anomalies and holography,''
arXiv:0802.1017 [hep-th].
}

\lref\BerensteinIJ{
D.~Berenstein, R.~Corrado, W.~Fischler and J.~M.~Maldacena,
``The operator product expansion for Wilson loops and surfaces in the large
$N$ limit,''
Phys.\ Rev.\ D {\bf 59}, 105023 (1999)
[arXiv:hep-th/9809188].
}

\lref\GiombiDE{
S.~Giombi, R.~Ricci and D.~Trancanelli,
``Operator product expansion of higher rank Wilson loops from D-branes and
matrix models,''
JHEP {\bf 0610}, 045 (2006)
[arXiv:hep-th/0608077].
}

\lref\BeasleyDC{
  C.~Beasley, J.~J.~Heckman and C.~Vafa,
  ``GUTs and Exceptional Branes in \break
  F-theory - I,''
  arXiv:0802.3391 [hep-th].
}

\lref\SemenoffAM{
G.~W.~Semenoff and D.~Young,
``Exact $1/4$ BPS loop - Chiral primary correlator,''
Phys.\ Lett.\ B {\bf 643}, 195 (2006)
[arXiv:hep-th/0609158].
}

\lref\ChenZZR{
B.~Chen, C.~Y.~Liu and J.~B.~Wu,
``Operator product expansion of Wilson surfaces from M5-branes,''
JHEP {\bf 0801}, 007 (2008)
[arXiv:0711.2194 [hep-th]].
}

\lref\CorradoPI{
  R.~Corrado, B.~Florea and R.~McNees,
  ``Correlation functions of operators and Wilson surfaces in the
  $d = 6$, $(0,2)$
  theory in the large $N$ limit,''
  Phys.\ Rev.\  D {\bf 60}, 085011 (1999)
  [arXiv:hep-th/9902153].
}

\lref\ReyIK{
  S.~J.~Rey and J.~T.~Yee,
``Macroscopic strings as heavy quarks in large $N$ gauge theory and
anti-de Sitter supergravity,''
  Eur.\ Phys.\ J.\  C {\bf 22}, 379 (2001)
  [arXiv:hep-th/9803001].
}

\lref\MaldacenaIM{
  J.~M.~Maldacena,
``Wilson loops in large $N$ field theories,''
  Phys.\ Rev.\ Lett.\  {\bf 80}, 4859 (1998)
  [arXiv:hep-th/9803002].
}

\lref\DrukkerZQ{
N.~Drukker, D.~J.~Gross and H.~Ooguri,
``Wilson loops and minimal surfaces,''
Phys.\ Rev.\  D {\bf 60}, 125006 (1999)
[arXiv:hep-th/9904191].
}

\lref\EricksonAF{
J.~K.~Erickson, G.~W.~Semenoff and K.~Zarembo,
``Wilson loops in $\cN = 4$ supersymmetric Yang-Mills theory,''
Nucl.\ Phys.\  B {\bf 582}, 155 (2000)
[arXiv:hep-th/0003055].
}

\lref\DrukkerRR{
N.~Drukker and D.~J.~Gross,
``An exact prediction of $\cN = 4$ SUSYM theory for string theory,''
J.\ Math.\ Phys.\  {\bf 42}, 2896 (2001)
[arXiv:hep-th/0010274].
}

\lref\DrukkerKX{
N.~Drukker and B.~Fiol,
``All-genus calculation of Wilson loops using D-branes,''
JHEP {\bf 0502}, 010 (2005)
[arXiv:hep-th/0501109].
}

\lref\GomisIM{
J.~Gomis and F.~Passerini,
``Wilson loops as D3-branes,''
JHEP {\bf 0701}, 097 (2007)
[arXiv:hep-th/0612022].
}

\lref\HartnollIS{
S.~A.~Hartnoll and S.~P.~Kumar,
``Higher rank Wilson loops from a matrix model,''
JHEP {\bf 0608}, 026 (2006)
[arXiv:hep-th/0605027].
}

\lref\OkuyamaJC{
K.~Okuyama and G.~W.~Semenoff,
``Wilson loops in $\cN = 4$ SYM and fermion droplets,''
JHEP {\bf 0606}, 057 (2006)
[arXiv:hep-th/0604209].
}

\lref\LuninXR{
O.~Lunin,
``On gravitational description of Wilson lines,''
JHEP {\bf 0606}, 026 (2006)
[arXiv:hep-th/0604133].
}

\lref\GomisSB{
J.~Gomis and F.~Passerini,
``Holographic Wilson loops,''
JHEP {\bf 0608}, 074 (2006)
[arXiv:hep-th/0604007].
}

\lref\YamaguchiTQ{
S.~Yamaguchi,
``Wilson loops of anti-symmetric representation and D5-branes,''
JHEP {\bf 0605}, 037 (2006)
[arXiv:hep-th/0603208].
}

\lref\HartnollHR{
S.~A.~Hartnoll and S.~Prem Kumar,
``Multiply wound Polyakov loops at strong coupling,''
Phys.\ Rev.\  D {\bf 74}, 026001 (2006)
[arXiv:hep-th/0603190].
}

\lref\YamaguchiTE{
S.~Yamaguchi,
``Bubbling geometries for half BPS Wilson lines,''
Int.\ J.\ Mod.\ Phys.\  A {\bf 22}, 1353 (2007)
[arXiv:hep-th/0601089].
}

\lref\DHokerFQ{
E.~D'Hoker, J.~Estes and M.~Gutperle,
``Gravity duals of half-BPS Wilson loops,''
JHEP {\bf 0706}, 063 (2007)
[arXiv:0705.1004 [hep-th]].
}

\lref\SemenoffXP{
G.~W.~Semenoff and K.~Zarembo,
``More exact predictions of SUSYM for string theory,''
Nucl.\ Phys.\  B {\bf 616}, 34 (2001)
[arXiv:hep-th/0106015].
}

\lref\PestunRZ{
V.~Pestun,
``Localization of gauge theory on a four-sphere and supersymmetric Wilson
loops,''
arXiv:0712.2824 [hep-th].
}

\lref\LuninAB{
O.~Lunin,
``$1/2$-BPS states in M theory and defects in the dual CFTs,''
JHEP {\bf 0710}, 014 (2007)
[arXiv:0704.3442 [hep-th]].
}

\lref\Fefferman{
C.~Fefferman and R.~Graham,
``Conformal Invariants,'' Ast\`erisque, hors s\'erie,
1995, p.95.
}

\lref\ChenIR{
B.~Chen, W.~He, J.~B.~Wu and L.~Zhang,
``M5-branes and Wilson Surfaces,''
JHEP {\bf 0708}, 067 (2007)
[arXiv:0707.3978 [hep-th]].
}

\lref\SkenderisVF{
  K.~Skenderis and M.~Taylor,
  ``Branes in $AdS$ and pp-wave spacetimes,''
  JHEP {\bf 0206}, 025 (2002)
  [arXiv:hep-th/0204054].
}

\lref\HenningsonXI{
  M.~Henningson and K.~Skenderis,
  ``Weyl anomaly for Wilson surfaces,''
  JHEP {\bf 9906}, 012 (1999)
  [arXiv:hep-th/9905163].
}
\lref\GustavssonHN{
  A.~Gustavsson,
  ``On the Weyl anomaly of Wilson surfaces,''
  JHEP {\bf 0312}, 059 (2003)
  [arXiv:hep-th/0310037].
}
\lref\GustavssonGJ{
  A.~Gustavsson,
  ``Conformal anomaly of Wilson surface observables: A field theoretical
  computation,''
  JHEP {\bf 0407}, 074 (2004)
  [arXiv:hep-th/0404150].
}

\lref\HofmanXT{
  D.~M.~Hofman and J.~M.~Maldacena,
``Giant magnons,''
J.\ Phys.\ A  {\bf 39}, 13095 (2006)
[arXiv:hep-th/0604135].
}

\lref\DrukkerGA{
N.~Drukker,
``$1/4$ BPS circular loops, unstable worldsheet instantons and the matrix
model,''
JHEP {\bf 0609}, 004 (2006)
[arXiv:hep-th/0605151].
}

\lref\DrukkerCU{
N.~Drukker and B.~Fiol,
``On the integrability of Wilson loops in $AdS_5\times S^5$: Some
periodic ansatze,''
JHEP {\bf 0601}, 056 (2006)
[arXiv:hep-th/0506058].
}

\lref\BuchbinderAR{
  E.~I.~Buchbinder, J.~Gomis and F.~Passerini,
  ``Holographic Gauge Theories in Background Fields and Surface Operators,''
  JHEP {\bf 0712}, 101 (2007)
  [arXiv:0710.5170 [hep-th]].
}

\lref\HarveyAB{
  J.~A.~Harvey and A.~B.~Royston,
  ``Localized Modes at a D-brane--O-plane Intersection and Heterotic Alice
  Strings,''
  JHEP {\bf 0804}, 018 (2008)
  [arXiv:0709.1482 [hep-th]].
}

\Title{
\vbox{\baselineskip12pt
\hbox{HU-EP-08/17}
\hbox{arXiv:0805.4199}}}
{\vbox{\centerline{Probing $\cN=4$ SYM With Surface Operators}}}

\vskip-5pt
\centerline{Nadav Drukker$^{1a}$, Jaume Gomis$^{2b}$ and
Shunji Matsuura$^{2,3c}$}

\smallskip
\smallskip
\bigskip
\centerline{\it$^{1}$Humboldt-Universit\"at zu Berlin, Institut f\"ur Physik}
\centerline{\it Newtonstra$\beta$e 15, D-12489 Berlin, Germany}
\medskip
\medskip
\centerline{\it$^2$Perimeter Institute for Theoretical Physics}
\centerline{\it Waterloo, Ontario N2L 2Y5, Canada}
\medskip
\medskip
\centerline{\it$^3$Department of Physics}
\centerline{\it University of Tokyo, 7-3-1 Hongo, Tokyo}
\footnote{${}^{}$}{${}^{a}$\tt drukker@physik.hu-berlin.de}
\footnote{${}^{}$}{${}^{b}$\tt jgomis@perimeterinstitute.ca}
\footnote{${}^{}$}{${}^{c}$\tt smatsuura@perimeterinstitute.ca}
\vskip .1in
\centerline{\bf{Abstract}}
\smallskip
\noindent
 ~~~~~In this paper we study  surface operators  in $\cN=4$ supersymmetric  Yang-Mills theory.\break
 We compute surface operator observables,  such as 
 the expectation value of  surface operators, the 
correlation function of surface operators with local operators,  and  the 
correlation function of surface operators with Wilson and 't Hooft loops. 
The calculations are performed using three different realizations of surface operators, corresponding respectively to 
the gauge theory path integral definition, the probe brane description  in $AdS_5\times S^5$  and the  ``bubbling''  supergravity 
 description of surface operators.
We find remarkable agreement between   the different calculations performed 
using the three different  realizations.

\Date{05/2008}
\listtoc\writetoc

\newsec{Introduction and Summary}

In gauge theories, nonlocal operators can be understood as operators that insert a probe into the theory.
Well known  are the Wilson and 't Hooft operators, which are characterized geometrically by a curve in spacetime. A Wilson loop inserts into the theory a point-like charged particle while an 't Hooft operator inserts a magnetic monopole. These operators play an important role  as order parameters of various phases of gauge theory. 

A surface operator on the other hand is characterized geometrically by a surface $\Sigma$ in spacetime. This surface  corresponds  to the worldsheet spanned by a probe string that the surface operator inserts in spacetime. 
A surface operator $\ope$ measures the response of the external string with which the field theory is probed, and  therefore these operators may enlarge the list of known order parameters  in gauge theories. 

In ${\cal N}=4$ SYM two types of supersymmetric surface operators have been studied. One is akin to the 't Hooft operator in that it is an operator of disorder type, characterized by  a certain supersymmetric vortex-like singularity  near a surface $\Sigma$ \GukovJK.
One can also construct a supersymmetric surface  operator of order type \BuchbinderAR, akin to a Wilson loop operator, which inserts into the ${\cal N}=4$ SYM path integral the  WZW action integrated over $\Sigma$ together with an associated Chern-Simons term, where the WZW action is constructed out of  the ${\cal N}=4$ SYM gauge field along $\Sigma$ (see also the   work in \HarveyAB)\foot{Disorder surface operators with higher pole singularities have been constructed in 
\lref\WittenTD{
  E.~Witten,
  ``Gauge Theory And Wild Ramification,''
  arXiv:0710.0631 [hep-th].
}
\WittenTD\ 
while rigid disorder surface operators have been studied in  
\lref\GukovSN{
  S.~Gukov and E.~Witten,
  ``Rigid Surface Operators,''
  arXiv:0804.1561 [hep-th].
}
\GukovSN. Disorder surface operators have also been considered in \BeasleyDC.}. 

In this paper we go beyond identifying and classifying these operators and 
perform three classes of computations with the surface operators \op. 
We  first study properties associated to
the surface operator \op\ itself, namely evaluating its expectation value. 
Then we consider the interaction of a surface operator \op\  
with local operators, including chiral primary operators and the stress-energy tensor. And then we analyze the 
interaction between a surface operator \op\ and   other non-local operators in 
the gauge theory --- certain supersymmetric Wilson loops and 't Hooft loops.
These correlators exhibit a rich dependence on the parameters that a surface 
operator \op\  is characterized by, which we review in Section 2.1.

We study the disorder surface operators $\ope$ of \GukovJK\ by using three 
alternative realizations of them. In the first one we use the gauge 
theory definition in terms of supersymmetric  codimension two singularities in 
the $\cN=4$ SYM path integral \GukovJK. In the second,  a surface operator  
\op\ is realized in terms of a configuration of D3-branes  ending on the 
boundary of \ads\ along the surface $\Sigma$, which realizes  the boundary 
conditions corresponding to a surface operator \op\ in  the gauge theory living 
on the boundary. The third description of surface operators \op\  is in terms of ``bubbling'' geometries --- 
ten dimensional solutions  of Type IIB supergravity which are asymptotically 
$AdS_5\times S^5$. They were proposed in 
\GomisFI\ as the gravitational description of surface operators and are related to 
the solutions of Lin, Lunin and Maldacena (LLM) \LinNB\LinNH\ by analytic  continuation.

While these correlation functions are computed using a variety of techniques, 
ranging from purely field theoretical to string theoretic calculations 
using strings, branes and supergravity solutions, the results are very coherent.
A rather complete picture arises that allows us to 
study the properties of surface operators  in ${\cal N}=4$ SYM in different 
regimes of the gauge theory parameter space. It is quite remarkable that the 
results of our computations computed in various regimes of $\cN=4$ SYM show 
  agreement with each other. The computations performed using semiclassical 
techniques in the gauge theory agree with the computations we perform using a bulk 
dual string theory computation.

The  three realizations of a surface operator \op\ are valid in different regimes of 
the gauge theory parameter space. The semiclassical gauge theory description is 
appropriate at weak 't Hooft coupling $\lambda\ll 1$. The probe D-brane description 
of a surface operator is valid when $N\gg1$, $\lambda \gg 1$ and when the number of 
D3-branes making up the corresponding surface operator is small. Finally, the 
supergravity description in terms of ``bubbling" geometries is appropriate when 
$N\gg1$, $\lambda \gg 1$ and when the surface operator \op\ is     characterized
in the probe description by a large number of D3-branes, in which case the backreaction of the D3 branes cannot be neglected. Nevertheless, surprisingly, 
we obtain quantitative agreement  when the   the same computation involving surface 
operators is performed using the different realizations of \op.

\medskip 
\noindent
{\it Summary of Results}
\smallskip

Before getting into the details of the calculations in the following
sections, let us elaborate a bit on the quantities we compute in this paper.

 The first quantity we calculate is the expectation value of the surface
operator \op. In this paper we consider maximally supersymmetric  surface
operators \op, which can have the geometry of the plane or of a
sphere, that is $\Sigma=R^2$ or $S^2$. Naively, one may think that in a conformal
field theory like $\cN=4$ SYM, that the expectation value of \op\  cannot depend on the radius $a$ 
 of the $S^2$ on which \op\ is supported. However, just as  for local operators, surface operators
can  have conformal anomalies which introduce a non-trivial scale-dependence.
Conformal invariance, however,  restricts the functional form of the expectation value to be
\eqn\restexp{
\vev{\cO}=C\,a^\Delta\,,
}
where in principle both $C$ and $\Delta$ can   depend on the coupling
constant of the theory\foot{In the case of local operators one can determine the dimension
by calculating a 2-point function. We would like to point out, to
avoid confusion, that the analogous quantity here is the vacuum expectation value of 
a {\it single} sphere, not two. In both cases there is a single length-scale,
the distance between the points or the radius.}.
We have calculated   these quantities in the semiclassical approximation of the gauge theory and also in the 
  probe approximation, where a surface operator \op\ is described by a configuration of D3 branes. In both    regimes we find that
\eqn\cdelta{
C=1,\qquad\Delta=0\,.
}
This result is rather different from the one obtained for surface observables in the six-dimensional
$(0,2)$ theory dual to string theory on $AdS_7\times S^4$, where the
dimension $\Delta$ is non-zero \BerensteinIJ\GrahamW\HenningsonXI\GustavssonHN\GustavssonGJ.

We then calculate the correlation function between a surface operator \op\ and local operators  $\cO$ in $\cN=4$ SYM. 
We are able to calculate these  correlation functions using the three different realizations of surface operators, one using gauge theory, one using probe D3-branes and one using the ``bubbling" supergravity solutions.
We find  remarkable 
 agreement between these correlators computed in the three different regimes.

Such two-point functions are obvious quantities to study
since local operators serve as natural probes  of surface operators \op, and
 it is interesting to unravel how they interact with them.
 Moreover, when the surface operator \op\ is supported on an $S^2$,  it can be approximated 
 at large distances by a linear
combination of local operators \BerensteinIJ. In other words, there is an
operator product expansion of a surface operator  \op\ in terms of local
ones $\cO$. The correlators of the surface operator \op\ with the local operators $\cO$ 
determine the coefficients in this expansion.

One particularly interesting local operator is the energy-momentum
tensor $T_{mn}$. The correlator between $T_{mn}$ and a local operator $\cO$ is
proportional to the conformal dimension of the operator $\cO$.  When we calculate the 
correlator between $T_{mn}$ and a surface operator \op\ we find a non-zero answer. This is not in
contradiction with  the already stated result that there is no radius $a$ dependence in the
vacuum expectation value of a surface operator \op. As we summarize in 
in Section~2.2,  a surface operator \op\ has    three different possible conformal
anomalies  and the two-point function of a  surface
operator with the energy-momentum tensor may calculate a different combination 
of these three anomaly coefficients.

We then calculate the correlation function between a surface operator \op\ and Wilson and t'Hooft loop operators in $\cN=4$ SYM. 
We are also able to calculate these  correlation functions using the three different realizations of surface operators, one using gauge theory, one using probe D3-branes and one using the ``bubbling" supergravity solutions.
We also find  remarkable agreement between these correlators computed in the three different regimes.

The geometrical properties of a correlator between a surface operator \op\ supported on a surface $\Sigma$  in spacetime and a loop operator supported on a curve $C$ in spacetime are quite interesting, since  in four dimensions  a surface $\Sigma$ can be linked by a curve $C$. The correlators of \op\ with Wilson and 't Hooft loops  exhibit   rich dependence  on all the parameters that a surface operator \op\ is characterized by.

The Wilson loop we consider was constructed in \DrukkerCU\ and preserves eight 
supercharges. This Wilson loop links the surface operator \op\ and depends on two 
parameters $(\theta_0,\psi_0)$ which encode  the coupling of the six scalars in 
$\cN=4$ SYM to the loop. It has been argued in  \DrukkerCU\  that the expectation 
value of this Wilson loop  is captured by a matrix model. We expect a similar matrix 
model to be valid also 
in the case when a surface operator \op\ is inserted into the path integral. This 
suggests how to extrapolate the value of the correlator between \op\ and 
$W_{\theta_0,\psi_0}$ to strong coupling by including the strong coupling answer of 
the matrix model.

In the realization of the surface operator \op\ in terms of a probe D3-brane, we calculate the correlator between \op\ and $W_{\theta_0,\psi_0}$ by evaluating the on-shell action of a string worldsheet that goes between the boundary of \ads\ and which ends on the probe D3-brane in the bulk, and we show that it reproduces the matrix model result.

In the realization of the surface operator \op\ in terms of a ``bubbling" supergravity solution, we describe the Wilson loop $W_{\theta_0,\psi_0}$ by  classical worldsheets in the exact ``bubbling" supergravity background.
 Remarkably, the string worldsheet is described by straight line  in the auxilliary space of the ``bubbling" geometry. The correlator between \op\ and $W_{\theta_0,\psi_0}$ is then obtained by summing over all string worldsheets that are saddle points of the path integral,   reproducing the matrix model result.
 
 Analogous calculations are performed for a supersymmetric 't Hooft loop $T_{\theta_0,\psi_0}$, but in this case there is no known matrix model description of the operator and the gauge theory analysis is performed in the semiclassical regime.

The paper is
organized according to the different realizations of a surface operator \op, and in each realization we perform 
several calculations. In Section $2$ we consider the gauge theory description of \op, where we study conformal anomalies, the vacuum expectation value of \op, as well as the correlators of \op\ with local and 
Wilson and 't~Hooft
loop operators $W_{\theta_0,\psi_0}$ and  $T_{\theta_0,\psi_0}$. 

In Section $3$ we study surface operators and their correlation functions using a probe D3-brane description. 
 Here we evaluate the classical
on-shell action of the D3-brane  and find the overall scaling anomaly as well as   the
correlator of \op\ with chiral primary operators in $\cN=4$ SYM. We also  find the
string solution  describing the Wilson loop $W_{\theta_0,\psi_0}$ in the presence of a surface operator \op\  and calculate the the correlator 
between the two operators. We extend this analysis to the case when the Wilson loop is replaced by the 't Hooft loop $T_{\theta_0,\psi_0}$.

In Section 4 we turn to the exact supergravity ``bubbling" solutions describing  
surface operators. From the exact supergravity solution we find the correlator of \op\ 
with various local operators $\cO$ using holographic renormalization. We then find the exact description of the Wilson loop $W_{\theta_0,\psi_0}$ 
in terms of a classical worldsheet in the full supergravity background corresponding to a surface operator \op. 
We show that these worldsheets have a very simple description in the ``bubbling" solution and calculate the correlator between \op\ and  $W_{\theta_0,\psi_0}$  by evaluating the on-shell action of the string worldsheet. A similar analysis is considered for the 't Hooft loop $T_{\theta_0,\psi_0}$

We end with a discussion where, in particular,  we
 speculate about a possible matrix model dual of a surface operator \op. Some computations are relegated to the appendices.

\newsec{Gauge Theory Description}

In this section we perform calculations  with   surface operators in
$\cN=4$ SYM in the semiclassical approximation
of the gauge theory.
 We start with a brief review of  disorder surface operators \GukovJK\ (see also \GomisFI) and their semiclassical description. We then calculate their expectation value,
 the correlators  of  a surface operator  with local operators and the correlators of  a surface operator  with a
Wilson and an 't~Hooft loop.

\subsec{Surface Operators in ${\cal N}=4$ SYM}

\lref\RohmXK{
  R.~M.~Rohm,
  ``Some Current Problems In Particle Physics Beyond The Standard Model,''
}

\lref\PreskillBM{
  J.~Preskill and L.~M.~Krauss,
  ``Local Discrete Symmetry And Quantum Mechanical Hair,''
  Nucl.\ Phys.\  B {\bf 341}, 50 (1990).
}

A disorder  surface operator $\ope$ supported on a surface
$\Sigma \subset R^4$ is characterized by a codimension two
singularity for the $\cN=4$ SYM fields\foot{For previous work
involving codimension two singularities in gauge theory see
\RohmXK\PreskillBM.}. In this paper we are interested in
maximally supersymmetric  surface operators in ${\cal N}=4$ SYM which preserve the
maximal subgroup%
\foot{This is the supergroup for a maximally supersymmetric surface operator in $R^{1,3}$.
In this paper we consider the theory in $R^4$, but surface  operators
exist in both.} 
$PSU(1,1|2)\times PSU(1,1|2)\times U(1)\subset
PSU(2,2|4)$ of the four dimensional  superconformal group
of ${\cal N}=4$ SYM.
The corresponding Euclidean supergroup has an $SO(1,3)\times SO(2)\times SO(4)$ bosonic
subgroup,
of which the first component acts on space-time alone, the last
is an $R$-symmetry and the $SO(2)$ is a diagonal combination of
space-time and $R$-symmetries. $SO(1,3)$ is
the Euclidean conformal group in two dimensions and the only 2-dimensional
submanifolds of
$R^4$ which posses this symmetry are $R^2$ and $S^2$, which will
be the locus of support of our surface operators%
\foot{The  two surfaces are related to each other by a global conformal
transformation. On $R^{1,3}$ there are also hyperboloids with geometry
$H_2$ and $dS_2$ which are conformal to space-like and time-like
planes respectively. One should be able to construct maximally
supersymmetric surface operators on them too.}.

The symmetries allow a complex scalar field in
the ${\cal N}=4$ SYM multiplet to develop  a pole near $\Sigma$,
reducing the $SO(1,5)\times SU(4)$ symmetry of ${\cal N}=4$
SYM to
$SO(1,3)\times SO(4)\times U(1)$.

 An alternative way to study  a surface operator $\ope$ is to consider the gauge theory on\foot{Throughout this paper  $AdS$ refers always to Euclidean $AdS$ space.} $AdS_3\times S^1$ instead of $R^4$,   the two geometries being related by a Weyl transformation. Then a surface operator $\ope$ is characterized by  a non-trivial boundary condition for the fields in the gauge theory near the boundary of $AdS_3\times S^1$. The advantage of this formulation is that the symmetries of the surface operator  $\ope$ are made manifest, as  they are realized as isometries of $AdS_3\times S^1$ instead of as conformal symmetries in  $R^4$. The choice of surface $\Sigma$ is now captured by the topology of the $AdS_3\times S^1$ conformal boundary.
  For $\Sigma=S^2$ one
must consider the $AdS_3$ metric in global coordinates while for
$\Sigma=R^2$ one must consider the metric in Poincar\'e coordinates.

A surface operator $\ope$ in $R^4$ induces  a non-abelian
vortex configuration for the gauge field near the surface operator \op, which produces a codimension two singularity at $\Sigma$. The gauge field configuration breaks the
gauge group $G$ to a subgroup $L\subset G$ near $\Sigma$. The
singularity in the gauge field corresponding to a surface operator
$\ope$ with $L = \prod_{l=1}^M U(N_l)\subset U(N)$
is given by
\eqn\holo{
A= \pmatrix{\alpha_{1}\otimes 1_{N_1}&0 &\cdots & 0\cr
0& \alpha_{2}\otimes1_{N_2}&\cdots & 0 \cr
\vdots &\vdots &\ddots&\vdots\cr
0&0&\cdots&\alpha_{M}\otimes1_{N_M}}d\psi,}
where $\psi$ is the polar angle in the local normal bundle to $\Sigma$
  and $1_n$ is the $n$-dimensional unit matrix. The parameters $\alpha_i$ take values in the maximal torus $T^N=R^N/Z^N$ of the gauge group $G=U(N)$.

The operator $\ope$ also introduces into the path integral
two-dimensional $\theta$-angles that couple to the unbroken $U(1)$'s
along $\Sigma$. This can be represented by the operator insertion
\eqn\insertion{
\exp\left(i\sum_{l=1}^M\eta_l\int_\Sigma\hbox{Tr}\;
{F^{(l)}}\right),}
where
$F^{(l)}$ is the field-strength for each of the unbroken $U(1)$ factors. The parameters $\eta_i$ take values in the maximal torus of the $S$-dual
or Langlands dual gauge group $^{L}G$ \GukovJK.
Since the Langlands dual of $G=U(N)$ is also $^LG=U(N)$, we have that the
matrix of $\theta$-angles of a surface operator \op\ is also given by an
$L$-invariant matrix 
\eqn\angle{
\eta= \pmatrix{{\eta_{1}}\otimes 1_{N_1}&0 &\cdots & 0\cr
0& {\eta_{2}}\otimes1_{N_2}&\cdots & 0 \cr
\vdots &\vdots &\ddots&\vdots\cr
0&0&\cdots&{\eta_{M}}\otimes1_{N_M}}.}

A  maximally supersymmetric surface operator $\ope$  also excites
a complex scalar field in $\cN=4$ SYM which we label by
$\Phi\equiv{1\over \sqrt{2}}(\phi^5+i\phi^6)$. It develops an $L$-invariant
pole near $\Sigma$ of the form
\eqn\pole{
\Phi={1\over \sqrt{2}\,z}\pmatrix{\beta_{1}+i\gamma_1\otimes 1_{N_1}
&0 &\cdots & 0\cr
0& \beta_{2}+i\gamma_2\otimes1_{N_2}&\cdots & 0 \cr
\vdots &\vdots &\ddots&\vdots\cr
0&0&\cdots&\beta_{M}+i\gamma_M\otimes1_{N_M}},}
where $z=r\,e^{i\psi}$ is the complex coordinate in the local normal bundle to $\Sigma$.
This form of the singularity is  completely determined by imposing superconformal invariance and
$L$-invariance.

In summary, a maximally supersymmetric surface operator \op\ in ${\cal N}=4$ SYM
with Levi group $L=\prod_{l=1}^M U(N_l)$ is labeled by $4M$
$L$-invariant parameters $(\alpha_l,\beta_l,\gamma_l,\eta_l)$.
The surface operator in $R^4$ is defined by performing the $\cN=4$ SYM
path integral over smooth field fluctuations around the $L$-invariant
singularities \holo\pole\ with the insertion of the operator \insertion.
We must mod out the path integral by
gauge transformations that take values in $L\subset U(N)$ when
restricted to $\Sigma$. The surface operator \op\ becomes singular
whenever the parameters that label the surface operator
$(\alpha_l,\beta_l,\gamma_l,\eta_l)$ for $l=1,\cdots, M$ are such
that they are invariant under a larger symmetry than $L$, the group of
gauge transformations we have to mode out by when evaluating the path
integral.

In the $AdS_3\times S^1$ picture, the surface operator $\ope$
is described by a non-trivial holonomy of the gauge field \holo\ around the non-contractible $S^1$ parametrized by the   coordinate $\psi$ and by the insertion of \insertion, where $\Sigma$ is now the conformal boundary of $AdS_3\times S^1$. The operator $\ope$ is also characterized by a non-singular configuration for the scalar field
\eqn\nopole{
\Phi={e^{-i\psi}\over\sqrt2}\pmatrix{\beta_{1}+i\gamma_1\otimes 1_{N_1}
&0 &\cdots & 0\cr
0& \beta_{2}+i\gamma_2\otimes1_{N_2}&\cdots & 0 \cr
\vdots &\vdots &\ddots&\vdots\cr
0&0&\cdots&\beta_{M}+i\gamma_M\otimes1_{N_M}},}
where $\psi$ is the coordinate parametrizing the $S^1$ in $AdS_3\times S^1$. Therefore, a  surface operator $\ope$ is  described by a half BPS vacuum state of the gauge theory on $AdS_3\times S^1$.

Note that the field configurations for $\Sigma=R^2$ and for $S^2$ are
exactly the same when the gauge theory is on $AdS_3\times S^1$.
For $\cN=4$ SYM on $R^4$ on the other hand, the description of
the singularities of the fields produced by
a surface operator with $\Sigma=S^2$ of radius $a$ is quite different
from $\Sigma=R^2$. One can perform a Weyl transformation from
global $AdS_3\times S^1$ to $R^4$ and use that the gauge field
and the scalar fields transform as weight zero and weight one
fields respectively. Starting with the metric of $R^4$ in spherical coordinates
\eqn\Rfourspherical{
ds^2=dr^2+r^2(d\vartheta^2+\sin^2\vartheta\,d\varphi^2)+dx^2\,,
}
and defining
\eqn\rtilde{
\tilde r^2={(r^2+x^2-a^2)^2+4a^2x^2\over 4a^2}
={R^2\over(\cosh\rho-\cos\psi)^2}\,,\qquad
r=\tilde r\sinh\rho\,,\qquad
x=\tilde r\sin\psi\,,
}
we find the metric
\eqn\RfourAdSS{
ds^2=\tilde r^2\left(d\rho^2+\sinh^2\rho\,
(d\vartheta^2+\sin^2\vartheta\,d\varphi^2)
+d\psi^2\right)\,,
}
which is conformal to  $AdS_3\times S^1$ in global coordinates.

In the  coordinate system \Rfourspherical, a surface operator \op\ with $\Sigma=S^2\subset R^4$ located at $r=a$ produces a singularity on the scalar fields given by \pole\ with the
 replacement of $z\rightarrow \tilde r\,e^{i\psi}$, where $\psi$ is written in terms of the
  $R^4$ coordinates   \Rfourspherical\ using \rtilde.
The gauge field singularity is exactly the same as before \holo,
where again
$d\psi$ is written in terms of  the $R^4$ coordinates    \Rfourspherical\
using \rtilde. Finally, the insertion of the $\theta$-angles \insertion\ is
now obtained by integrating $\hbox{Tr} F_l$ over the $S^2$ parametrized by
$\vartheta$ and $\varphi$ in \Rfourspherical\ at $r=a$.

\subsec{Surface Operator Expectation Value}

We now compute the expectation value of the surface operator $\ope$
in the  semiclassical approximation of the gauge theory. We want
to calculate the expectation value for both $\Sigma=R^2$
and $\Sigma=S^2$. When the operator is supported on $R^2$ we
anticipate that the expectation value will be one by supersymmetry.
For the case $\Sigma=S^2$ the expectation value may
depend non-trivially on the radius $a$ of the $S^2$.

Since we are dealing with operators defined on a surface
$\Sigma$, an even dimensional submanifold in spacetime, the
surface operator $\ope$ may posses conformal anomalies
\BerensteinIJ\GrahamW. The conformal anomalies for surface operators are given by local expressions that depend on the pullback onto the surface $\Sigma$ of the curvature invariants in spacetime and also on the extrinsic curvature, which encodes the embedding of the surface $\Sigma$ in spacetime.

The classification of the surface conformal anomalies can be rephrased as a cohomology problem, as in the case of usual conformal anomalies
\lref\BonoraCQ{
  L.~Bonora, P.~Pasti and M.~Bregola,
  ``Weyl Cocycles,''
  Class.\ Quant.\ Grav.\  {\bf 3}, 635 (1986).
}
\BonoraCQ. One must classify solutions to the Wess-Zumino consistency conditions modulo terms that can be obtained as the Weyl variation of a local term. A basis of the Weyl cohomology has been recently presented in
\SchwimmerYH\ (for related previous work on surface anomalies in six dimensions see
\HenningsonXI\GustavssonHN\GustavssonGJ). The surface anomaly
depends on three anomaly coefficient $c_i$, $i=1,2,3$
\eqn\anom{
{\cal A}_\Sigma={1\over 4\pi}\int_\Sigma d^2\sigma
\sqrt{h}\left(c_1R^{(2)}+c_2g_{mn}\left(h^{\mu\sigma}h^{\nu\rho}
-{1\over 2}h^{\mu\nu}h^{\rho\sigma}\right)
K_{\mu\nu}^m K_{\rho\sigma}^n+c_3h^{\mu\nu}h^{\rho\sigma}C_{\mu\rho\nu\sigma}\right),}
where $h_{\mu\nu}$ and $g_{mn}$ denote the induced metric on $\Sigma$ and the spacetime metric respectively while $C_{mnpq}$ and $K^m_{\mu\nu}$ denote the Weyl tensor and the extrinsic curvature. The anomaly coefficients $c_i$ are the analogs of the well known $a$ and $c$ anomaly coefficients in four dimensions.
The expression for  $c_i$ depend on the choice of field theory as well as on the choice of surface operator in that field theory.

For the case  $\Sigma=R^2\subset R^4$ we have that     ${\cal A}_{R^2}=0$. For $\Sigma=S^2\subset R^4$ one gets 
\eqn\intanom{
{1\over4\pi}\int_\Sigma \sqrt{h}\,R^{(2)}=2\,,\qquad
g_{mn}
\left(h^{\mu\sigma}h^{\nu\rho}
-{1\over2}h^{\mu\nu}h^{\rho\sigma}\right)
K^m_{\mu\nu}K^n_{\rho\sigma}=C_{\mu\rho\nu\sigma}=0.
}
Therefore ${\cal A}_{S^2}=2c_1$.

Under a conformal transformation $g_{mn}\to e^{2\sigma}g_{mn}$
the expectation value of a surface operator $\ope$ supported
on $S^2\subset R^4$ changes by an amount proportional to the anomaly 
\eqn\change{
\delta_{\sigma}\vev{\ope}=2c_1\vev{\ope}.}
 Therefore the expectation value of $\ope$ is determined in terms of the radius $a$ of the $S^2$  by 
\eqn\scaling{
\vev{\ope}\propto a^{2c_1}.
}
The precise form of the expectation value of $\ope$ now depends on the chosen operator and on the theory under discussion, which uniquely determine the anomaly coefficients $c_i$.

We now focus on our specific case of the maximally supersymmetric surface 
operators $\ope$ in $\cN=4$ SYM and calculate their vacuum expectation value 
in the semiclassical approximation. This is achieved by
evaluating the classical SYM
action on the field configuration produced by the operator $\ope$
\eqn\vevex{
\vev{\ope}= \exp(-\cS)|_{\rm surface}.}
It is easiest to
perform  this calculation by considering the field configuration
produced by a surface operator $\ope$  on $AdS_3\times S^1$.
The relevant part of the action of $\cN=4$ SYM on
$AdS_3\times S^1$ is given by
\eqn\cladsact{
\cL={1\over  g_{YM}^2}\Tr\left(|D\Phi|^2+{R^{(4)}\over6}|\Phi|^2\right),
}
where ${R^{(4)}\over 6}|\Phi|^2$ is the conformal coupling for the
scalars and  $R^{(4)}=-6$ is the scalar curvature of $AdS_3\times S^1$.
The surface operator $\ope$ induces a non-trivial field configuration
on the field $\Phi$ \nopole\ on  $AdS_3\times S^1$ which depends only on the
angular coordinate $\psi$ of the
$S^1$ through $\Phi=\Phi_0 e^{-i\psi}$, where $\Phi_0$ is a constant
diagonal matrix \nopole. The operator $\ope$ also induces a nontrivial gauge field configuration \holo\  which nevertheless commutes with  $\Phi_0$, so that $D\Phi=\partial \Phi$.
Therefore $|D\Phi|^2=|\Phi|^2$ and
the two terms in the Lagrangian \cladsact\ exactly cancel each, so that 
\eqn\vevfinal{
\vev{\ope}=1.}
This result holds regardless of whether we use the
Poincar\'e patch or global $AdS_3$ metric in the field theory, so that
$\vev{\ope}=1$ for both  $\Sigma=R^2$ and $\Sigma=S^2$.

The same result can also be derived by studying $\cN=4$ SYM on $R^4$.
In this case the scalars do not have a mass term. However,
 one must  include in the action a boundary term along
the surface operator \op\ that guaranties that the boundary terms
in the equations of motion vanish. To see that, consider the
relevant part of the action in flat space
\eqn\bndrybulk{
\cS={1\over g_{YM}^2}\int d^4x\sqrt{g}\,g^{mn}\,
\Tr\left(D_m \Phi\,D_n\bar \Phi\right)}
with the metric \RfourAdSS.
In deriving the equations of motion there is a boundary term near
the locus of the surface operator at a cutoff $\rho=\rho_0$
\eqn\bndryeom{
\delta\cS={1\over g_{YM}^2}\int d\vartheta\,d\varphi\,d\psi\,
\tilde r^2\sinh^2\rho_0\sin\vartheta
\left(\delta \Phi\,\partial_\rho\bar \Phi
+\delta\bar \Phi\,\partial_\rho\Phi\right)\,.
}
This  does not vanish, since $\Phi$ diverges.
To fix this we need to add to the action the boundary term
\eqn\bndryterm{
\cS_{\rm boundary}=-{1\over  g_{YM}^2}\int d\vartheta\,d\varphi\,d\psi\,
{1\over a}\,\tilde r^3\sinh^3\rho_0\sin\vartheta\,
\Tr\left(\Phi\bar\Phi\right)\,.
}
The variation of the bulk and boundary action would now give the
boundary equation $\partial_\rho\Phi-\tilde r\sinh\rho_0\,\Phi/a=0$,
which is indeed satisfied by our solution.

Evaluating the classical action with a cutoff $\rho_0$ gives the
divergent result
\eqn\bulkclassaction{
\cS={1\over g_{YM}^2}\int d\rho\,d\psi\,d\vartheta\,d\varphi\,
{\cosh\rho\sinh^2\rho\sin\vartheta\over\cosh\rho-\cos\psi}
\sum_{l=1}^MN_l(\beta_l^2+\gamma_l^2)
=4\pi^2\sinh^2\rho_0\sum_{l=1}^MN_l(\beta_l^2+\gamma_l^2)\,.
}
The boundary term \bndryterm, though, exactly cancels this bulk
piece leading to a total  action that vanishes, exactly as
we found in the $AdS_3\times S^1$ picture.

It follows from \scaling\ and from  $\vev{\ope}=1$ for a spherical surface operator $\ope$ in $\cN=4$ SYM that  the anomaly coefficient  $c_1$ for this theory (and this family of surface operators) is $c_1=0$. It would be interesting to calculate separately the anomaly coefficients $ c_2,c_3$, which can be done by computing the anomaly for more general  surfaces $\Sigma$ as well as by considering $\cN=4$ SYM in more general curved four dimensional backgrounds.

\subsec{Correlator of Surface Operators with Local Operators}

We now proceed to calculate the correlation function of a surface operator $\ope$ with various local operators in $\cN=4$ SYM in $R^4$. We consider the correlator with the   dimension $\Delta=2,3$ chiral primary operators (CPO's)  ${\cal O}_\Delta$
of ${\cal N}=4$ SYM, the traceless stress-energy tensor $T_{mn}$ and with the $U(1)$ spacetime symmetry current $J_m^\psi$ which generates rotations in the space transverse to the surface $\Sigma$.

In the semiclassical approximation, the correlation function of a surface operator $\ope$ with a local operator ${\cal O}$ is obtained by evaluating  ${\cal O}$ in the background field produced by the surface operator $\ope$
\eqn\opeshellII{{\vev{\ope\cdot{\cal O}}\over \vev{\ope}}
= {\cal O}|_{\rm surface}.}

If instead we consider the surface operator $\ope$ as a non-trivial boundary condition for $\cN=4$ SYM  on $AdS_3\times S^1$,
the correlation function of $\ope$ with a local operator ${\cal O}$ in $R^4$ corresponds to the vacuum expectation value of the
local operator ${\cal O}$ in the state $\left|\ope \right>$.
Thus on $AdS_3\times S^1$ the correlation function \opeshellII\ in $R^4$ corresponds to 
\eqn\vacuumexpect{
\vev{\cO^I}_{\ope}.
}
In all these computations the position dependence of the local operator is determined by conformal
Ward-Takahashi identities. In the case of $\Sigma=R^2\subset R^4$ the
correlator of a surface operator $\ope$ with a local operator ${\cal O}$ of dimension $\Delta$ scales with the distance $r$ between the local operator ${\cal O}$ and the plane where the surface operator $\ope$ is supported by%
\foot{For operators that are not scalars, the tensor structure of the correlator is determined by Lorentz invariance.}
\eqn\scaleopeR{
{\vev{\cO\cdot\ope}\over \vev{\ope}}={C_{\cO}\over r^{2\Delta}}\,.
}
The quantity that we need to compute for the various local operators $\cO$ is $C_{\cO}$.

When the surface operator $\ope$ is supported on $S^2\subset R^4$,
the correlator is given by
\eqn\scaleopeRnew{
{\vev{\cO\cdot\ope}\over \vev{\ope}}={C_{\cO}\over \tilde{r}^{2\Delta}}\,.
}
where $\tilde r$ is the conformally invariant distance \rtilde
\eqn\radial{
\tilde r={\sqrt{(r^2+x^2-a^2)^2+4a^2x^2}\over2a}\,,}
combining $a$, the radius of the $S^2$, the radial position of the local
operator in the $R^3$ including the surface, $r$ and the transverse
distance to the local operator, $x$.
 In principle the proportionality constant
in \scaleopeR\ \scaleopeRnew\ $C_{\cO}$ should be the same for $\Sigma=S^2$ and $\Sigma=R^2$,
except that there is the possibility that some contributions that exist for the
sphere are ``pushed to infinity'' in the case of the plane and will not show up.
This does not happen in the specific calculations we performed but may
show up once one includes quantum corrections.

When one considers the gauge theory on $AdS_3\times S^1$
the surface operator $\ope$ is supported
at the boundary of $AdS_3$. Therefore, there
is no invariant distance like $r$ or $\tilde r$ in $R^4$. In
this case
\eqn\vacuumexpecteq{
\vev{\cO}_{\ope}=C_{\cO}\,.
}

The CPO's are scalar operators transforming in the $(0,\Delta,0)$ representation of the $SU(4)$ R-symmetry. The operators are given by
\eqn\unit{
{\cal O}^I_\Delta={(8\pi^2)^{\Delta/2}\over \lambda^{\Delta/2} \sqrt{\Delta}}C^I_{i_1\ldots i_\Delta}\hbox{Tr}\left(\phi^{i_1}\ldots\phi^{i_\Delta}\right),}
where $\lambda=g_{YM}^2N$ is the 't Hooft coupling and
$Y^{I}=C^{I}_{i_1\ldots i_\Delta}x^{i_1}\ldots x^{i_\Delta}$
are the $SO(6)$ scalar spherical harmonics. The operators \unit\ are normalized such that 
their two point
function is unit normalized
\lref\LeeBXA{
  S.~Lee, S.~Minwalla, M.~Rangamani and N.~Seiberg,
  ``Three-point functions of chiral operators in $D = 4$, $\cN = 4$ SYM at  large
  $N$,''
  Adv.\ Theor.\ Math.\ Phys.\  {\bf 2}, 697 (1998)
  [arXiv:hep-th/9806074].
}
\eqn\twopoint{
\vev{{\cal O}^I_\Delta(x){\cal O}^J_\Delta(y)}={\delta^{IJ}\over |x-y|^{2\Delta}}.}

A surface operator $\ope$   preserves an $SO(4)$ subgroup of the
$SU(4)$ R-symmetry group. Therefore, the correlation function
of a chiral primary operator   ${\cal O}^I_\Delta$ with a surface operator
$\ope$ is   nontrivial only  if the CPO is an $SO(4)$ singlet.
By looking at the decomposition of the $(0,\Delta,0)$ representation
of $SO(6)\simeq SU(4)$ under the obvious $SO(4)\times SO(2)$
subgroup one finds that there are
$\Delta+1$ CPO's of dimension $\Delta$ which are singlets under
$SO(4)$. These $SO(4)$ invariant operators   have charges $k=-\Delta,\,-\Delta+2,\cdots,\Delta-2,\Delta$
under $SO(2)$.
We label the $SO(4)$ invariant spherical harmonics by
$Y^{\Delta,k}=C^{\Delta,k}_{i_1\ldots i_\Delta}x^{i_1}
\ldots x^{i_\Delta}$ (for more details see
Appendix~A).
With these   spherical harmonics we can write down the $SO(4)$ invariant CPO's as
\eqn\unitproj{
{\cal O}_{\Delta,k}={(8\pi^2)^{\Delta/2}
\over \lambda^{\Delta/2} \sqrt{\Delta}}C^{\Delta,k}_{i_1\ldots i_\Delta}\hbox{Tr}\left(\phi^{i_1}\ldots\phi^{i_\Delta}\right).}

Using the explicit expression for the $SO(4)$ invariant
spherical harmonics with $\Delta=2,3$, we can write down
the unit normalized CPO's
\eqn\opers{\eqalign{
{\cal O}_{2,0}&={4\pi^2\over \sqrt{6}\lambda}
\hbox{Tr}\left(4\Phi\bar{\Phi}-\sum_{I=1}^4\phi^I\phi^I\right);\hskip.65in
{\cal O}_{2,2}={8\pi^2\over \sqrt{2}\lambda}\hbox{Tr}\left(\Phi^2\right);
\cr
{\cal O}_{3,1}&={8\pi^3\over \lambda^{3/2}}
\hbox{Tr}\left(2\Phi^2\bar{\Phi}-\Phi\sum_{I=1}^4\phi^I\phi^I\right);\qquad
{\cal O}_{3,3}={32\pi^3\over\sqrt{6}\lambda^{3/2}}
\hbox{Tr}\left(\Phi^3\right),
}}
  where as before $\Phi={1\over \sqrt{2}}(\phi^5+i\phi^6)$ 
and $\cO_{\Delta,-k}={\bar \cO}_{\Delta,k}$.

The semiclassical correlations function \opeshellII\ of these CPO's with the surface operator $\ope$ are then given by 
\eqn\correlafin{\eqalign{
{\vev{\cO_{2,0}\cdot\ope}\over \vev{\ope}}&={1\over |z|^2} {8\pi^2\over \sqrt{6}\lambda}\sum_{l=1}^MN_l (\beta_l^2+\gamma_l^2);~~~~~~
{\vev{\cO_{2,2}\cdot\ope}\over \vev{\ope}}={1\over z^2}{4\pi^2\over \sqrt{2}\lambda}\sum_{l=1}^MN_l (\beta_l+i\gamma_l)^2\cr
{\vev{\cO_{3,1}\cdot\ope}\over \vev{\ope}}&={1\over z|z|^2}
{ 8\pi^3\over \sqrt{2}\lambda^{3/2}}\sum_{l=1}^MN_l (\beta_l^2+\gamma^2_l)(\beta_l+i\gamma_l);\cr
{\vev{\cO_{3,3}\cdot\ope}\over \vev{\ope}}&={1\over z^3}
{8\pi^3\over\sqrt{3}\lambda^{3/2}}\sum_{l=1}^MN_l (\beta_l+i\gamma_l)^3\,.}}
The remaining correlators  can be obtained  by complex conjugating \correlafin.

These result are purely classical and they may receive 
quantum corrections. The surface operator \op\  breaks near $\Sigma$ 
 the gauge group from $U(N)$ to
$L = \prod_{l=1}^M U(N_l)$, and also  breaks $R$-symmetry group down to
$SO(4)$. This  breaking may introduce different interactions 
among the different fields at the quantum level. We therefore expect 
the quantum corrections to depend on the choice of Levi group given 
by the rank of the unbroken gauge groups  $N_1,\cdots ,N_M$. 
The calculations performed using probe D3-branes and the ``bubbling" supergravity solution in the next two sections 
indicate, however, that the loop corrections terminate after a finite number of loops! The 
perturbative corrections to the correlator of \op\ with $\cO_{\Delta,k}$  terminate at
order\foot{Note that $\Delta-|k|$ is always  even, so we get a standard perturbation expansion in the 't Hooft coupling $\lambda$.}  $\lambda^{(\Delta-|k|)/2}$. It would be   interesting to understand the truncation of the loop 
corrections directly in the gauge theory.

\smallskip

We now consider the correlator of $\ope$ with the stress-energy tensor $T_{mn}$ of $\cN=4$ SYM in $R^4$, which was already computed in \GomisFI. In the  presence of a planar surface operator in $R^4$ the correlator is fixed up to function $h$ 
\eqn\oper{
{\vev{T_{\mu\nu}\cdot\ope}\over\vev{\ope}}
=h{\eta_{\mu\nu}\over r^4},
\qquad
{\vev{T_{ij}\cdot\ope}\over\vev{\ope}}
={h\over r^4}\left[{4n_in_j-3\delta_{ij}}\right],
\qquad
\vev{T_{\mu i}\cdot\ope}=0\,.
}
Here $x^{m}=(x^\mu,x^i)$, where $x^\mu$ are coordinates along
$\Sigma=R^2$ and $n^i=x^i/r$ is the unit normal vector to the planar surface. If $\cN=4$ SYM is on $AdS_3\times S^1$ the expectation value of the  stress-energy tensor $T_{mn}$ in the presence of a surface operator \op\ is given by
\eqn\correlaadss{
\vev{T_{ab}}_{\ope} =h\;g_{ab}  \qquad\qquad
\vev{T_{\psi\psi}}_{\ope}=-3h,} 
where $g_{ab}$ is the metric on $AdS_3$ and $\psi$ is the coordinate on the $S^1$.

The piece of the  stress-energy tensor of $\cN=4$ SYM which involves the gauge field and the complex scalar field $\Phi$ excited by the surface operator $\ope$ is given by(see e.g.
\lref\GubserSE{
  S.~S.~Gubser and I.~R.~Klebanov,
  ``Absorption by branes and Schwinger terms in the world volume theory,''
  Phys.\ Lett.\  B {\bf 413}, 41 (1997)
  [arXiv:hep-th/9708005].
}
\GubserSE)\foot{We neglect the scalar potential term, as it does not contribute to the correlator.}
\eqn\emten{\eqalign{
T_{mn}=&{2\over 3g^{2}_{YM}}\Tr\Big[2D_{m}\Phi D_{n}\bar\Phi
+2D_{m}\bar\Phi D_{n}\Phi-g_{mn}|D\Phi|^2
-\Phi D_{m}D_{n}\bar{\Phi}-\bar{\Phi} D_{m}D_{n}\Phi \cr
&+{1\over 4}g_{mn}\Phi D^2\bar{\Phi}+{1\over 4}g_{mn}\bar{\Phi} D^2 \Phi \Big]
+{2\over g^{2}_{YM}}
\Tr\left[-F_{ml}F_{nl}+{1\over4}g_{mn}F_{lp}F_{lp}\right].
}}
Using the semiclassical formula \opeshellII\ we find that 
\eqn\dimen{
h=-{2\over 3g_{YM}^2}{\sum_{l=1}^MN_l(\beta_l^2+\gamma_l^2)}\,.
}
or equivalently
\eqn\dimen{\eqalign{
\vev{T_{ab}}_{\ope}&=-{2\over 3g^2_{YM}}{\sum_{l=1}^MN_l(\beta_l^2+\gamma_l^2)}g_{ab}\cr
\vev{T_{\psi\psi}}_{\ope}&={2\over  g^2_{YM}}{\sum_{l=1}^MN_l(\beta_l^2+\gamma_l^2)}\,.}}
 In this case, we expect that the only perturbative correction appears at the one loop level, of order $\lambda$, which  as we will see is captured by the ``bubbling" supergravity solution.

In \KapustinPY\ the analogous coefficient $h$ for Wilson and 't Hooft
loop operators was dubbed the scaling weight,
  the name suggesting  that it should generalize the notion of conformal
dimension of local conformal fields to extended objects. The analogy with
conformal dimension of a local operators is not complete. The conformal
dimension of a local operator measures its behavior under conformal
transformations, while for surface operators there are   three different
anomalies under such transformations which we dubbed
$c_i$ in \anom. We expect the scaling weight $h$ to be related   to the anomaly coefficients
$c_i$ and it would be interesting to find the explicit relation.
\smallskip
We also calculate the correlator of \op\ with the conserved
current $J_m^{\psi}$. This current generates the $U(1)$ symmetry that acts by shifts on the
$S^1$ of the $AdS_3\times S^1$ geometry on which the gauge theory is defined.
However, it is not a symmetry of the surface operator \op\
as the singularity induced on  the scalar field \pole\ is not $U(1)$ invariant.
 In conformal field theories, the current associated with a
spacetime symmetry is constructed by contracting the stress-energy
tensor $T_{mn}$ with the conformal Killing vector $\xi_G$ generating
the symmetry as $j_{m}^G=T_{mn}\xi_G^n$. Therefore, the correlator  of
$J_m^\psi$ with \op\ is given by
\eqn\opecurre{
\vev{J_\psi^\psi}_{\ope}=  \vev{T_{\psi\psi}}_{\ope}={2\over  g^2_{YM}}{\sum_{l=1}^MN_l(\beta_l^2+\gamma_l^2)}\,.}

\subsec{Correlator with Wilson and 't  Hooft Loop Operator}

The last calculation we perform is the correlator of a surface
operator \op\ with a Wilson loop operator
and also make comments about  the correlator of a surface operator \op\ with an 't Hooft loop operator.
  This combination of operators is particularly
interesting since in four dimensional space a two dimensional surface
can be  linked by a curve.

We   take the Wilson loop supported on a curve $C$ to link the surface $\Sigma$ on which the surface
operator \op\ is supported.
The simplest way to realize this configuration is by taking the
surface $\Sigma=R^2\subset R^4$
 and consider a circular loop $C=S^1$
in the orthogonal plane. A natural choice of Wilson loop to consider   is the $1/4$
BPS circular Wilson loop with periodic coupling to the scalar fields
with arbitrary phase shift $\psi_0$
\ZaremboWL\eqn\gaugeWLpsi{
W_{\psi_0}={1\over N}\hbox{Tr}\,\exp\int\left(
iA_\psi+r\cos(\psi-\psi_0)\,\phi^5-r\sin(\psi-\psi_0)\,\phi^6\right)d\psi\,.
}
Here $\phi^5$ and $\phi^6$ are the real and imaginary part of the 
complex field turned on by the surface operator \op,  given by
$\Phi\equiv{1\over \sqrt{2}}(\phi^5+i\phi^6)$.

We consider  an even more general Wilson loop operator, presented in
\DrukkerCU\ and studied further in \DrukkerGA, with an extra
coupling to a third real scalar field $\phi^1$ in the $\cN=4$ SYM multiplet, with a relative coupling
given by the angle $\theta_0$
\eqn\gaugeWL{
W_{\theta_0,\psi_0}
={1\over N}\,\hbox{Tr}\,\exp\int d\psi\left(
iA_\psi+|z|\cos\theta_0\,\phi^1
+\sqrt{2}\,\sin\theta_0\Re(z\Phi e^{-i\psi_0})\right).
}
Fort $\theta_0=\pi/2$ it reduces to \gaugeWLpsi\
(note that the $\psi$ dependence comes now from $z=re^{i\psi}$), while for
$\theta_0=0$ it's the well studied maximally supersymmetric circular loop
\EricksonAF\DrukkerRR.

Note that this Wilson loop operator depends on two parameters
$\theta_0$ and $\psi_0$ and operators with different
values of these parameters are quite different.
In particular, the supercharges that are preserved by the Wilson loop $W_{\theta_0,\psi_0}$
depend on these parameters, though they are always compatible with those
of the surface operator. This is proven in Appendix~B.

Evaluating the correlator of a surface operator \op\ with the  Wilson loop \gaugeWL\ in the semiclassical approximation  amounts to evaluating the Wilson loop operator \gaugeWL\ on the fields
 produced
by the surface operator  \holo\pole.
The result is given by
\eqn\gaugeWLvalueI{
{\vev{W_{\theta_0,\psi_0}\cdot\ope}_{\rm class}\over\vev{\ope}}
=\sum_{l=1}^M {N_l\over N}
\exp\Big[2\pi\sin\theta_0(\beta_l\cos\psi_0+\gamma_l\sin\psi_0)
+2\pi i\alpha_l)\Big]\,.
}
The extra scalar $\phi^1$ does not contribute at the classical level,
as the surface operator \op\ does not excite it.

In the absence of a surface operator \op, it is conjectured in \DrukkerGA\
that the exact expectation value of the Wilson loop $W_{\theta_0,\psi_0}$   is given
  by summing up all ladder diagrams in the   Feynman gauge, which is captured by
  a matrix model. In the limit of large 't Hooft coupling $\lambda$ it has 
an asymptotic expansion with two saddle points \DrukkerGA\ which can be written 
schematically as
  \eqn\vevlatWL{
\vev{W_{\theta_0,\psi_0}}\simeq
\exp\left[\sqrt\lambda\cos\theta_0\right]
+\exp\left[-\sqrt\lambda\cos\theta_0\right].
}
This result, including both saddle points can be reproduced from string theory. 

The expression \vevlatWL\ can be regarded as coming purely from the scalar 
field $\phi^1$, which
appears with a factor of $\cos\theta_0$ in the loop \gaugeWL, while
 the other two scalars  are present to  guarantee supersymmetry.

The presence of the surface operator can, of course, modify the contribution of $\phi^1$ 
to the expectation value of the Wilson loop \gaugeWL, but we now argue that its contribution 
is well under control. Indeed 
this is what happens for the Wilson loop  $W_{\theta_0,\psi_0}$   in the presence of a chiral primary 
local operator \SemenoffAM\ (see also \SemenoffXP\OkuyamaJC), where the exact result 
is given by a normal matrix model, but the asymptotics are still governed by an exponent 
like in \vevlatWL.

In particular, one possible effect of the surface operator \op\ on the scalar 
$\phi^1$ in the Wilson loop \gaugeWL\ is in breaking of the $U(N)$
gauge group to the Levi group $L=\prod_{l=1}^M U(N_l)$, thus  reducing the
number of degrees of freedom in the system. Now, in each $U(N_l)$ sector,  $\phi^1$
 gives the same matrix model result   with the replacement
$\lambda\to\lambda N_L/N$. Assuming that all the $N_l$
are still large, so that we may still use the approximation in \vevlatWL,
we predict that the correlator of the Wilson loop $W_{\theta_0,\psi_0}$  with a surface
operator \op\ at strong coupling is given by
\eqn\gaugeWLvalue{\eqalign{
{\vev{W_{\theta_0,\psi_0}\cdot\ope}\over\vev{\ope}}
=&\sum_{l=1}^M {N_l\over N}
\left(\exp\left[\sqrt{\lambda N_L\over N}\cos\theta_0\right]
+\exp\left[-\sqrt{\lambda N_L\over N}\cos\theta_0\right]\right)
\times\cr&\hskip1in{}\times
\exp\left[2\pi\sin\theta_0(\beta_l\cos\psi_0+\gamma_l\sin\psi_0)
+2\pi i\alpha_l)\right]\,.
}}
We will later see that this result is reproduced by performing a string theory calculation in the ``bubbling" supergravity background dual to a surface operator.

We note that unlike local operators, this class of Wilson loops
are sensitive to the Aharonov-Bohm phase \holo\ induced by the
surface operator \op.

Similarly to the Wilson loop, one can consider an 't~Hooft loop, which
should yield the S-dual result (accounting of course also for the
S-duality transformation of the surface operator parameters $(\alpha_l,\beta_l,\gamma_l,\eta_l)$). Unlike the Wilson loop, the 't Hooft loop does not have a simple
weak-coupling description. Rather, much like the surface operator \op, it
can be thought of as requiring certain singularities for the gauge field
and scalar field along the loop.

The 't Hooft loop $T_{\theta_0,\psi_0}$ we want to consider should be the dual of the Wilson
loop \gaugeWL, which couples to three scalars and depends on the
parameters $\theta_0$ and $\psi_0$.
An 't Hooft loop can be represented by a Dirac monopole embedded along   some
$U(1)$ subgroup of $U(N)$ with  the accompanying scalar source. The
field-strength around the monopole has the form
\eqn\Fthooft{
F={1\over 2}\,T^{0}\sin\theta\,d\theta\,d\phi\,,
}
where $\theta$ and $\phi$ are local coordinates on a sphere surrounding
the loop, and $T^0$ is a generator   in the Cartan subalgebra of $U(N)$.  This should be the behavior of the gauge field close to the 't Hooft loop  $T_{\theta_0,\psi_0}$ for the straight line. In the classical approximation this is the exact solution.
So consider now the spherical surface operator \op\ wrapping the straight
't Hooft loop $T_{\theta_0,\psi_0}$. The surface operator couples to the magnetic field produced by the 't Hooft operator through the
parameters $\eta$ in \insertion\ entering the definition of \op
\eqn\etathooft{
\exp\left(i\eta \int_\Sigma \hbox{Tr}\;
{T^0\over 2 }\right)=\exp\left(2\pi i \eta\right).
}
This produces the $S$-dual of the phase $\exp(2\pi i \alpha)$ in \gaugeWLvalue\ obtained from the correlator of \op\ with the  Wilson loop \gaugeWL, as $\alpha\rightarrow \eta$ under $S$-duality \GukovJK.

The contribution from the scalars $\Phi$ and $\bar\Phi$
is almost identical to the one in the Wilson loop calculation,
as the effect of the singularity on the scalar fields produced
by the 't Hooft loop can be modeled in the semiclassical
approximation by inserting into the $\cN=4$ SYM path integral
the operator
\eqn\insertloop{
\exp \left({4\pi\over g_{YM}^2} \int  \hbox{Tr}
\left( T^0 \Re(z\Phi e^{-i\psi_0})\right)d\psi\right).}
Inserting the field  \pole\ produced  by the surface operator
\op\ in \insertloop\ gives the $S$-dual of the Wilson loop
result  \gaugeWLvalueI\ as
$|\beta+i\gamma|\rightarrow 4\pi/g^2_{YM} |\beta+i\gamma|$
under $S$-duality \GukovJK. The contribution of $\phi^1$ as in
\vevlatWL\ has so far not been reproduced from the 't Hooft
operator in a gauge-theory calculation
even in the absence of the surface operator \op, and we
do not attempt to recover that here. We will see in Section~3 and in Section~4, where 
we calculate the Wilson loop and 't Hooft loop using string theory, that the resulting 
expressions are indeed related to each-other by $S$-duality.

We note that the correlator of $\ope$ with an 't Hooft operator is sensitive to the   two dimensional $\theta$-angles \insertion\ that appear in the definition of a surface operator \op.

\newsec{Probe Description in $AdS_5\times S^5$}

In this section we perform calculations  with   surface operators in
$\cN=4$ SYM in the probe approximation of the dual Type IIB string theory on $AdS_5\times S^5$.
 We start with a description   of surface operators in ${\cal N}=4$ SYM in terms of D3-branes in $AdS_5\times S^5$. We then calculate their expectation value,
 the correlators  of  a surface operator  with CPO's and the correlators of   a surface operator  with a
Wilson and an 't~Hooft loop.

\subsec{Surface Operator as D-branes in $AdS_5\times S^5$}

A surface operator $\ope$ in $\cN=4$ SYM corresponds in the probe
picture to having a D3-brane in $AdS_5\times S^5$ ending on the
boundary along the prescribed surface $\Sigma$. This D3-brane
solution in  $AdS_5\times S^5$  was first studied by \Constable.

We start by describing the details of the $AdS_3\times S^1$ D3-brane solution that we need to perform the various calculations in this section. We take the metric on the $S^5$ to be
\eqn\metricsphere{
d\Omega_5=\cos^2\theta d\Omega_3+d\theta^2+\sin^2\theta d\phi^2,}
which makes manifest the $SO(4)\times SO(2)\subset SO(6)$ symmetry of the $S^5$.
For the surface operator \op\ supported on $\Sigma=R^2$ it is convenient to write the bulk $AdS$ metric in two different slicings
\eqn\metplaneI{
ds^2_{(1)}={1\over y^2}\left(
dy^2+dx_0^2+dx_1^2+dr^2+r^2\,d\psi^2\right)\,,
}
\eqn\metplaneII{
~~~~~~~~~~~~~ds^2_{(2)}={\cosh^2u\over z^2}
(dz^2+dx_0^2+dx_1^2)+du^2+\sinh^2u\,d\psi^2,
}
where we work in units where the $AdS_5$ and $S^5$ radius of curvature is one so that $L^4=4\pi g_s N l_s^4=1$.
The two coordinate systems are related by 
\eqn\planecoords{
y={z\over\cosh u}\,,\qquad
r=z\tanh u\,.
}
The first choice of coordinates \metplaneI\ is more suitable  when
considering the dual $\cN=4$ SYM theory on $R^4$. The second one \metplaneII\ is
more suitable when
$\cN=4$ SYM  is on $AdS_3\times S^1$, where $AdS_3\times S^1$ is the metric on the conformal boundary located at  $u\rightarrow \infty$.

We choose the RR four-form potential   in the analog of the
Fefferman-Graham gauge \Fefferman\ for the metric, so that  the
four-form has no components transverse to the $AdS_5$ boundary.
Therefore, we take the RR four-form potential    in the metric
\metplaneI\ and \metplaneII\ to be given by 
\eqn\CplaneI{
C_4^{(1)}={r\over y^4}\, dx_0\wedge dx_1\wedge dr\wedge
d\psi\,, } \eqn\CplaneII{ C_4^{(2)}={\cosh^4u\over z^3}\, dz\wedge
dx_0\wedge dx_1\wedge d\psi\,.
}

The worldvolume coordinates of the D3-brane embedding in the
coordinate system \metplaneI\ are  given by $x_0$, $x_1$, $r$
and $\psi$, and the D3-brane has  non-trivial embedding
functions  $y=y(r)$ and on the $S^5$ in the coordinate system
 \metricsphere: $\theta=\pi/2$
and $\phi=\phi(\psi)$.
With this ansatz the D3-brane action is given by
\eqn\dbiplane{
\cS_{D3}=T_{D3}\int dx_0\,dx_1\,dr\,d\psi\,
{1\over y^4}
\sqrt{(1+y'^2)(r^2+y^2\dot\phi^2)}-T_{D3}\int dx_0\,dx_1\,dr\,d\psi\,
{r\over y^4}\,
}
where the D3-brane tension is given by
$
T_{D3} ={N\over 2\pi^2}$ (in the units where $L^4=1$).
The half-BPS solution is given by \Constable 
\eqn\yrsolution{
y(r)={1\over \kappa} \,r\,,\qquad
\psi+\phi=\phi_0\,,
}
where $\kappa$ and $\phi_0$ are integration constants.

In the coordinate system \metplaneII\ the D3-brane solution has
worldvolume coordinates $z$, $x^0$, $x^1$ and $\psi$ and the embedding
is given by $\theta=\pi/2$ in \metricsphere\ and
\eqn\constu{
\sinh u_0=\kappa\,,\qquad
\psi+\phi=\phi_0\,.
}
The induced metric on the brane is that of $AdS_3\times S^1$, so
that we have the freedom of turning on   a Wilson line for the
gauge field $A$ and dual gauge field $\tilde{A}$ living on the
D3-brane worldvolume along the non-contractible $S^1$. Therefore,
the D3-brane solution ending on $\Sigma=R^2$ on the boundary
depends on four  parameters
\GukovJK, which can be identified with those of the gauge theory
description of a surface operator $\ope$ via\foot{Note that in the units $L^4=1$, we can write alternatively
$\beta+i\gamma={1\over2\pi l_s^2}\sinh u_0\, e^{i\phi_0}\,$ (see also equation (4.9).}
 \eqn\defkap{\eqalign{
\alpha&=\oint {A\over 2\pi}\,,\cr
\beta+i\gamma&={\sqrt\lambda\over2\pi}\sinh u_0\, e^{i\phi_0}\,,\cr
\eta&=\oint {\tilde{A}\over 2\pi}\,.}}

Therefore, in the probe approximation a surface operator \op\
with unbroken gauge group $L=\prod_{l=1}^M U(N_L)$ and labeled
by the  $4M$ parameters $(\alpha_l,\beta_l,\gamma_l,\eta_l)$
corresponds to a configuration of $M$ stacks of D3 branes in
\ads\ ending on the boundary    on the surface $\Sigma$, with
each stack consisting of $N_l$ coincident D3 branes.
In Appendix~C we study the supersymmetry preserved by the probe D3-brane
brane and show that it is the same as the supersymmetry preserved by the surface operator \op\ in
$\cN=4$ SYM.

For the probe D3-brane description of the surface operator \op\ in $\cN=4$ SYM supported on $\Sigma=S^2$ it is convenient to write the bulk $AdS_5$ metric in two different slicings 
\eqn\metsphereI{
\hskip-92pt ds^2_{(3)}={1\over y^2}\left(
dy^2+dr^2+r^2(d\vartheta^2+\sin^2\vartheta\,d\varphi^2)
+dx^2\right)\,,
}
\eqn\metsphereII{
ds^2_{(4)}=\cosh^2u
(d\rho^2+\sinh^2\rho\,(d\vartheta^2+\sin^2\vartheta\,d\varphi^2))
+du^2+\sinh^2u\,d\psi^2.
}
The two coordinate systems are related  by
\eqn\spherey{
y={a\over\cosh u\cosh\rho-\sinh u\cos\psi}\,,
\qquad
r=y\cosh u\sinh\rho\,,
\qquad
x=y\sinh u\sin\psi,
}
where $a$ is the radius of the $\Sigma=S^2$ on which the dual surface operator $\ope$ is supported. Again, the  choice of coordinates \metsphereI\ is appropriate when
considering the gauge theory on $R^4$ while the one in  \metsphereII\
is appropriate when  the gauge theory is on $AdS_3\times S^1$.

The associated expressions for the RR four-form potential are  given by:
\eqn\CsphereI{
\hskip-60ptC_4^{(3)}={r^2\sin\vartheta\over y^4}\,
dr\wedge d\vartheta\wedge d\varphi\wedge dx\,,
}
\eqn\CsphereII{
C_4^{(4)}=\cosh^4u\sinh^2\rho\sin\psi\,
d\rho\wedge d\vartheta\wedge d\varphi\wedge d\psi\,.
}

The D3-brane worldvolume  coordinates  in the metric \metsphereII\
are $\rho$, $\vartheta$, $\varphi$ and $\psi$, and the  embedding is
again given by $\theta=\pi/2$ in \metricsphere\ and:
\eqn\spheresolu{ \sinh u_0=\kappa\,,\qquad \psi+\phi=\phi_0\,. } In
the coordinate system \metsphereI\ the D3-brane embedding
translates into
\eqn\spheretilde{
{(r^2+x^2+y^2-a^2)^2+4a^2x^2\over4a^2y^2}=\kappa^2\,.
}
The map between the parameters of the D3-brane solution and the
parameters defining the dual gauge theory surface operator $\ope$ is
also given by \defkap. As explained earlier, considering $M$ stacks of  D3 branes
yields a surface operator \op\ with parameters $(\alpha_l,\beta_l,\gamma_l,\eta_l)$  for $l=1,\ldots,M$.

\subsec{Expectation Value in the Probe Approximation}

The expectation value of the surface operator \op\ in
$\cN=4$ SYM corresponding to a single probe D3-brane is given
in the leading $N>>1$ and $\lambda >>1$ approximation by 
\eqn\vevprobe{
\vev{\ope}=e^{-\cS_{D3}|_{\rm on-shell}}.}
This corresponds to a surface operator \op\ with unbroken gauge
group $L=U(1)\times SU(N-1)$
and with parameters $(\alpha,\beta,\gamma,\eta)$. Therefore, we have to evaluate the classical action of the D3-brane solutions in the previous subsection.

For the case of a surface operator $\ope$ on $\Sigma=R^2$, we find that the action of the dual  D3-brane in the metric
 \metplaneI\ is given by
 \eqn\actionplane{
\cS_{D3}={N\over 2\pi^2}\int dx_0\,dx_1\,dr\,d\psi\,{1\over y^4}
\left[\sqrt{(1+y'^2)(r^2+y^2\dot\phi^2)}-r\right],
}
where $\cS_{D3}=\cS_{DBI}-\cS_{WZ}$.
In order to evaluate the classical action it is not enough to plug in the solution
$\dot\phi=-1$ and $y=r/\kappa$ into the action. Examining the
variation of the action one finds a boundary term that needs to be cancelled.
That is due to the fact that $y'$ is a non-zero constant and to get
a good variational problem it is
necessary to subtract
\eqn\planeLT{
y'\,{\partial\cL\over \partial y'}
={N\over 2\pi^2}{y'^2\over y^4}
\sqrt{r^2+y^2\dot\phi^2\over1+y'^2}\,.
}
Therefore,  the total D3-brane action is 
\eqn\actionplaneLT{
\cS_{D3}={N\over 2\pi^2}\int dx_0\,dx_1\,dr\,d\psi\,{1\over y^4}
\left[\sqrt{r^2+y^2\dot\phi^2\over1+y'^2}-r\right].
}
Plugging in the D3-brane solution \yrsolution\ one sees that the integrand vanishes $\cS_{D3}|_{\rm on-shell}=0$.

It is illustrative to repeat the calculation of the D3-brane action in the second coordinate system
\metplaneII, where now the field $u=u_0$ is a constant, so the variational
principle for it is well defined. The classical on-shell action  for the D3-brane   is now given by 
\eqn\actionplaneLTu{
\cS_{D3}|_{on-shell}={N\over 2\pi^2}\int dx_0\,dx_1\,dz\,d\psi\,
{\cosh^2u_0\over z^3}
\left[\cosh u_0\sqrt{\sinh^2u_0+\dot\phi^2}-\cosh^2u_0\right]=0.
}
We conclude that the expectation value of  a surface operator \op\ supported on $\Sigma=R^2$ in the probe approximation is given by
\eqn\finprobe{
\vev{\ope}=1,}
which agrees with the semiclassical gauge theory computation \vevfinal.

For the case of a surface operator $\ope$ supported on $\Sigma=S^2$, we find that the action of the dual  D3-brane   in the metric
 \metsphereII\ is given by 
\eqn\actionsphere{
\cS_{D3}={N\over2\pi^2}\int d\rho\,d\vartheta\,d\varphi\,d\psi\,
\cosh^3u_0\sinh^2\rho\sin\vartheta\left[
\sqrt{\sinh^2u_0+\dot\phi^2}-\cosh u_0\right]\,.
}
In this case $u=u_0$ is also a constant and it satisfies the proper variational principle so after
setting $\dot\phi=-1$ we find again that the on-shell action also vanishes $\cS_{D3}=0$. Thus the expectation value of  a surface operator \op\ supported on $\Sigma=S^2$ in the probe approximation is given by
\eqn\finprobe{
\vev{\ope}=1,}
agreeing with the semiclassical gauge theory computation \vevfinal.

The conformal anomaly \anom\ for the surface operator $\ope$ with $\Sigma=R^2$ and $S^2$ vanishes in the probe approximation and due to \scaling\ we have that $c_1=0$. It would be interesting to consider more general D3-brane solutions that end on a boundary with a different metric and
on a general surface $\Sigma$. This would allows to compute   the other anomaly coefficients $c_i$ in \anom.

\subsec{Correlator with Local Operators in Probe Approximation}

We now compute in the probe approximation the correlation function
of a surface operator \op\ and the  CPO's
$\cO^I_\Delta$ of dimension $\Delta=2,3$ in $\cN=4$ SYM.
Analogous computations have been done for Wilson loops in \BerensteinIJ\GiombiDE\SemenoffAM\ and for Wilson surfaces in six dimensions in \ChenZZR\CorradoPI.

A CPO  $\cO^I_\Delta$ in $\cN=4$ SYM couples to the  five dimensional scalar
 fluctuation $s^I$ \WittenQJ. This fluctuation mode diagonalizes the linearized equations of motion of Type IIB supergravity expanded around $AdS_5\times S^5$.  It solves the Klein-Gordon equation in $AdS_5$ \Nieuwen
\eqn\sphharm{
\nabla_\mu\nabla^\mu\,s^I=\Delta(\Delta-4)s^I\,,
}
where $\Delta$ is the dimension of the dual CPO $\cO^I_\Delta$.

In the gauge theory (Section~2.3) we saw that the only chiral
primaries that have non-trivial
coupling to the surface operators are the $SO(4)$ singlets. This can
be seen here from  the fact that in the metric \metricsphere\ the brane
is at $\theta=\pi/2$, where all the other spherical harmonics vanish.
We can combine all the $s^I$ which couple to the probe brane
into one field on $AdS_5\times S^5$ as
\eqn\relevantss{
s=\sum_\Delta\,\sum_{k=-\Delta,-\Delta+2,\cdots}^\Delta
Y^{\Delta,k}\left(\theta={\pi\over2},\phi\right)
s^{\Delta,k},
}
where $Y^{\Delta,k}$ are the $SO(4)$ invariant spherical harmonics of
$SO(6)$ (see Appendix~A). When evaluated on the D3-brane they are
given by 
\eqn\spherafixed{
Y^{\Delta,k}\left(\theta={\pi\over2},\phi\right)
=C_{\Delta,k}\,e^{ik \phi}.}

The correlator of a surface operator \op\ with $\cO^I_\Delta$ is
captured  in the probe approximation by the   solution of the
linearized equations of motion of Type IIB supergravity for
$s^I$ in the presence of the D3-brane source. In order to
calculate this we first need the bulk to boundary propagator
in  $AdS_5$. Let us consider $\cO^I_\Delta$ at a position
$(x'_0,\,x'_1,\,r',\,\psi')$ on the boundary of $AdS_5$
in the coordinate system \metplaneI.
The propagator from the insertion point to a  point in the bulk
is   given by
\eqn\propcpoa{
G=c\,{y^\Delta\over
((x-x')^2+r^2+r'^2-2rr'\cos(\psi-\psi')+y^2)^\Delta\,},
}
where $c={\Delta+1\over 2^{2-\Delta/2} N\sqrt{\Delta}}$ is chosen
\BerensteinIJ\ such that the bulk computation of the two-point
function of the dual CPO
operator $\cO^I_\Delta$ is unit normalized as in \twopoint. We find
it  convenient to work in  the coordinate system \metplaneII\ where
$AdS_5$ is foliated by $AdS_3\times S^1$ slices. If we place
$\cO^I_\Delta$ at  $(x'_0=x'_1=0,\,\psi',\,z'=d)$ then the bulk to boundary propagator reads 
\eqn\propcpo{
G=c\,{z^\Delta\over
\cosh^\Delta u_0(x_0^2+x_1^2+z^2-2dz\tanh u_0\cos(\psi-\psi')+d^2)^\Delta}.}

While the bulk field $s^I$ has a simple propagator, it has rather complicated couplings
to the usual supergravity fields. In the bulk it is related to the metric
on $AdS_5$, the metric on $S^5$ and the four forms on $AdS_5$ and $S^5$.
Following the notations of \Nieuwen, the relevant fields sourced by $s^I$ at linear order are 
\eqn\components{
h^{AdS}_{\mu\nu}\,,\qquad
h^{S}_{\alpha\beta}\,,\qquad
a^{AdS}_{\mu\nu\rho\sigma}\,,\qquad
a^{S}_{\alpha\beta\gamma\delta}\,.
}
The fluctuation $s^I$ sources these supergravity fields and these in turn couple to the probe D3-brane.
To compute the one point function we need to expand the probe D3-brane action to linear order in the fluctuations \components. The linearized fluctuation contribution from the D3-brane DBI action is
\eqn\cpodbi{\eqalign{
\cL_{DBI}^{(1)}&=
{T_{D3}\over 2}\int \sqrt{\det (g)}\,g^{ab}
(\partial_a X^\mu\partial_bX^\nu\,h^{AdS}_{\mu\nu}
+\partial_a X^\alpha\partial_bX^\beta\,h^S_{\alpha\beta})
 \cr&=
{T_{D3}\over 2}\int dz\,dx_0\,dx_1\,d\psi\,
{\cosh^2u\over z^3}
\left(z^2h^{AdS}_{zz}+z^2h^{AdS}_{00}+z^2h^{AdS}_{11}+h^{AdS}_{\psi\psi}
+h^S_{\phi\phi}\right),
}}
where $g$ is the induced metric.

The linearized fluctuation contribution from the WZ term of the D3-brane is:
\eqn\cpowz{
\cL_{WZ}^{(1)}
=T_{D3}\int a^{AdS}_{\mu\nu\rho\sigma}
}
As can be seen from the expressions above, our probe D3-brane, which is
mainly along $AdS_5$, does not couple to linear order to
$a^S_{\alpha\beta\gamma\delta}$, the 4-form on $S^5$.

The last ingredient we need is the relation between the diagonal fluctuation $s^I$ and the
fluctuations in the basis \components. This  is obtained by solving the linearized Type IIB supergravity equations of motion in the presence of the source
$s=\sum_Is^IY^I$, where $Y^I$ is an $SO(6)$ spherical harmonic. The solution is given by
 \Nieuwen\LeeBXA
\eqn\eomprobe{\eqalign{
h^{AdS}_{\mu\nu}&=-{6\over 5}\Delta\,s\,g_{\mu\nu}
+{4\over \Delta+1}\nabla_{(\mu}\nabla_{\nu)}s\,,\cr
h^S_{\alpha\beta}&=2\Delta\,s\,g_{\alpha\beta}\,,\cr
a^{AdS}_{\mu\nu\rho\sigma}&=-4\epsilon_{\mu\nu\rho\sigma\eta}
\nabla^\eta s\,.
}}
where $\nabla_\mu$ is the usual covariant derivative and
$\nabla_{(\mu}\nabla_{\nu)}s$ denotes the symmetric traceless
piece of $\nabla_{\mu}\nabla_{\nu}s$.

Finally we have all the pieces in place to calculate the correlator.
We plug into \cpodbi\ and \cpowz\ the expressions for the
fluctuations \components\ in terms of $s$ using \eomprobe.
For $s$ we take the bulk to boundary
propagator \propcpo\ with a boundary source
\eqn\sourcesymm{
s_0=\sum_\Delta \ \sum_{k=-\Delta,-\Delta+2,\cdots}^\Delta
Y^{\Delta,k}\left(\theta={\pi\over2},\phi\right) s^{\Delta,k}_0,
}
The  correlator  of a surface operator \op\ with the $\cN=4$ SYM CPO
$\cO_{\Delta,k}$  in \unitproj\ is given by the
the derivative of the D3-brane action  with respect to the source $s^{\Delta,k}_0(x)$. It is given by
\eqn\probcpoa{\eqalign{
{\vev{\ope\cdot\cO_{\Delta,k}}\over \vev{\ope}}
&=-{2^{1+\Delta/2}(\Delta+1)\sqrt{\Delta}T_{D3} d^2
C_{\Delta,k}\over N\cosh^{\Delta}u_0} \int dz\,dx_0\,dx_1\,d\psi\,
\cr&\hskip1in {z^{\Delta-1}e^{ik(\phi_0-\psi)}\over
(x_0^2+x_1^2+z^2-2dz\tanh u_0\cos(\psi-\psi')+d^2)^{\Delta+2},}}
}
where we have used that on the  D3-brane solution \constu\
$\psi+\phi=\phi_0$, where $\phi_0$ is one of the parameters   characterizing the
D3-brane embedding. Integrating over $x_0$ and $x_1$ and using that $T_{D3}=N/2\pi^2$ (in the units $L^4=1$) yields 
\eqn\chiralpripartial{
{\vev{\ope\cdot\cO_{\Delta,k}}\over \vev{\ope}}
={2^{\Delta/2}\sqrt{\Delta} d^2C_{\Delta,k}\over \pi\cosh^\Delta u_0}
\int_0^\infty dz\,\int_0^{2\pi} d\psi\, {z^{\Delta-1}e^{ik(\phi_0-\psi)}\over
(z^2-2dz\tanh u_0\cos(\psi-\psi')+d^2)^{\Delta+1}}\,.
}
We make a change of variables from $z$ to $\zeta$ defined by
\eqn\zzeta{
{2dz\tanh u_0\over z^2+d^2}=\sqrt{1-{1\over\zeta^2}}\,.
}
As $z$ varies between zero and infinity, $\zeta$ goes from
one to $\cosh u_0$ (at $z=d$) and back to one. The resulting integral
is then simplified by the fact that only terms with a branch-cut
in $\zeta$ will contribute. This gives
\eqn\probcpoc{\eqalign{
{\vev{\ope\cdot\cO_{\Delta,k}}\over \vev{\ope}}
=&{\sqrt{\Delta}C_{\Delta,k}e^{ik(\phi_0-\psi')}
\over 2^{\Delta/2+1}\pi d^\Delta\sinh^\Delta u_0}
2 \int_1^{\cosh u_0} d\zeta{\zeta\sinh u_0
\over\sqrt{\cosh^2u_0-\zeta^2}} (\zeta^2-1)^{\Delta/2-1}\times
\cr&
\times \int_0^{2\pi} d\tilde\psi\,{e^{ik\tilde\psi}\over
\left(\zeta-\sqrt{\zeta^2-1}\,\cos\tilde\psi\right)^{\Delta+1}}\,.
}}
We use now the integral representation of the associated
Legendre polynomial\foot{We use a slightly modified definition,
$P_\Delta^{|k|}(x)={1\over 2^\Delta\Delta!}
(x^2-1)^{|k|/2}(d^{\Delta+|k|}(x^2-1)^\Delta/d^{\Delta+|k|}x)$
which is single valued for argument greater than one.} 
\eqn\poly{
\int_0^{2\pi} d\tilde\psi\,{e^{ik\tilde\psi}\over
\left(\zeta-\sqrt{\zeta^2-1}\,\cos\tilde\psi\right)^{\Delta+1}}
=2\pi{(\Delta-|k|)!\over \Delta!}P_\Delta^{|k|}(\zeta).}
After a further change of variables
\eqn\zetachi{
\sqrt{\zeta^2-1}=\sinh u_0\sin\chi\,,
}
the correlator \probcpoc\ reduces to the integral 
\eqn\probcpojchi{
{(\Delta-k)!\sqrt{\Delta}C_{\Delta,k}e^{ik(\phi_0-\psi')}
\over \Delta!2^{\Delta/2}d^\Delta}
\int_0^\pi d\chi\sin^{\Delta-1}\chi\,
P_\Delta^{|k|}\left(\sqrt{1+\sinh^2u_0\sin^2\chi}\right).
}

We can now extract the correlators of the surface operator \op\ with any CPO 
$\cO_{\Delta,k}$. We write down explicitly the result for $\Delta=2$ and $3$ 
which we also did in the gauge theory in Section~2 and do in the 
``bubbling'' geometry description in the next section. 
Using the spherical harmonics in Appendix~A we have that
$C_{2,0}=1/\sqrt{3}$,  $C_{2,\pm2}=1/2$, $C_{3,\pm1}=\sqrt{3}/4$ and
$C_{3,\pm3}=1/2\sqrt2$
while the relevant associated Legendre polynomials are given by
$P_2^0(x)=(3x^2-1)/2$, $P_2^2(x)=3(x^2-1)$,
$P_3^1(x)=3\sqrt{x^2-1}(5x^2-1)/2$
and $P_3^3(x)=15(x^2-1)^{3/2}$. Performing the integral in
\probcpojchi\ we conclude that the correlator of  \op\ with
the CPO's \opers\ with $\Delta=2$ and $3$ are given in the probe approximation by 
\eqn\probcpofin{\eqalign{
{\vev{\cO_{2,0}\cdot \ope}\over \vev{\ope}}
&={2\over\sqrt6}\,{\cosh^2u_0\over d^2}\,;\hskip1.2in
{\vev{\cO_{2,2}\cdot \ope}\over \vev{\ope}}
={1\over\sqrt2}\,{\sinh^2u_0\over d^2}\,e^{2i(\phi_0-\psi')}\cr
{\vev{\cO_{3,1}\cdot \ope}\over \vev{\ope}}
&={1\over\sqrt2}\,{\cosh^2u_0\sinh u_0\over d^3}\,e^{i(\phi_0-\psi')}\,;\qquad
{\vev{\cO_{3,3}\cdot \ope}\over \vev{\ope}}
={1\over\sqrt3}\,{\sinh^3u_0\over d^3}\,e^{3i(\phi_0-\psi')}\,.
}}
The remaining correlators can be obtained by complex conjugating \probcpofin.

Recall that the probe calculation is done with a single D3-brane and
is dual to the surface operator \op\ in the gauge theory where the
gauge group is broken to $U(1)\times SU(N-1)$. 
In order to compare to the gauge theory result \correlafin\
we take $N_1=1$ and $N_2=N-1$, we use the identification 
\defkap\ to relate $e^{i\phi_0}\sinh u_0$ to $\beta_1+i\gamma_1$, 
and take $\beta_2=\gamma_2=0$. Then with
$z=d\,e^{i\psi'}$, we find that \probcpofin\ agrees precisely 
with \correlafin\ for $\cO_{2,2}$ and $\cO_{3,\pm3}$. The other correlators in 
\probcpofin\ exactly reproduce the semiclassical result in \correlafin\
but deviate from the classical result of the gauge theory  
by a quantum correction, proportional to the 't Hooft coupling $\lambda$. The general
form of the correlator \probcpojchi\ for $\cO_{\Delta,k}$ computed in the probe approximation indicates that the quantum corrections to the semiclassical gauge theory computation truncate to order 
$\lambda^{(\Delta-|k|)/2}$, suggesting that the gauge theory description of these correlators may be captured exactly by a reduced matrix model, which has a finite number of loop corrections.

\subsec{Correlator with Wilson Loop Operator}

We wish to examine the correlator of the surface operator \op\ with the
Wilson loop operator \gaugeWL. We employ the target-space metric
\metsphereII\ where the brane is at fixed $u=u_0$.
The Wilson loop is described, as usual, by a semiclassical string
surface \ReyIK\MaldacenaIM, ending on the boundary of \ads\ along
a path related to the specific choice of loop.
The specific loops we are considering are those studied in
\DrukkerCU\DrukkerGA, where the relevant boundary conditions are
further explained.
The string worldsheet is not at fixed $u$, and it varies from
$\infty$ to $u=u_0$, where it ends on the D3-brane probe.
In  the $u\to\infty$ region the
string worldsheet  ends in the \ads\ boundary along the desired
loop, which is a circle
along the $\psi$ direction at fixed\foot{
Due to conformal invariance the radius of the
loop in the gauge theory, and the value of $\rho$ here are
inconsequential}
$\rho$.
At the same time the string  also wraps the direction $\phi$
along a circle of $S^5$ at $\theta=\theta_0$ in the metric
\metricsphere.
The extra parameter $\psi_0$ which appeared in the definition of
the loop is the relative phase between the two angles, so that
at the boundary $\psi+\phi=\psi_0$,  matching the phase of the Wilson loop operator   in \gaugeWL.

Finding the string solution for this configuration is by now a standard
procedure. With the worldsheet coordinates $\sigma$ and $\tau$ one
assumes the rotationally-symmetric ansatz
\eqn\rotsymansatz{
u=  u(\sigma)\,,\qquad
\psi=\tau+\alpha(\sigma)\,,\qquad
\theta=\theta(\sigma)\,,\qquad
\phi=-\tau-\zeta(\sigma)\,,
}
and one can consistently keep all the other fields constant.
The boundary conditions at $u\rightarrow \infty$ on $\psi$ and
$\phi$ are such that $\alpha(0)-\zeta(0)=\psi_0$, so that at the boundary we get $\psi+\phi=\psi_0$, which is the phase corresponding to the Wilson loop \gaugeWL.
These extra two degrees of freedom are required in order to allow  the relative
phase of $\psi$ and $\phi$ to change as the string moves in the $u$ direction. At the other end of the
string worldsheet, where the string ends on the probe
D3-brane, we have that  $\psi+\phi=\phi_0$ \yrsolution, so
$\alpha-\zeta=\phi_0$ there.

The string Lagrangian in the conformal gauge is
\eqn\stringaction{
\cL= {\sqrt\lambda\over4\pi}
\big( u'^2+\sinh^2u\,(1+\alpha'^2)+\theta'^2
+\sin^2\theta\,(1+\zeta'^2)\big)\,.
}
Two conserved quantities are the momenta conjugate to $\alpha$ and
$\zeta$
\eqn\palphaeta{
p_\alpha=\sinh^2u\,\alpha'\,,\qquad
p_\zeta=\sin^2\theta\,\zeta'\,.
}
There are two Virasoro constraints
\eqn\virasoro{
\sinh^2u\,\alpha'+\sin^2\theta\,\zeta'=0\,.
\qquad
u'^2-\sinh^2u\,(1-\alpha'^2)
+\theta'^2-\sin^2\theta\,(1-\zeta'^2)=0\,.
}
The first one sets $p_\alpha=-p_\zeta$. By solving the equations
of motion for $u$ and $\theta$ one finds expressions similar to
the second Virasoro constraint for each independently
\eqn\eomprobeWL{
u'^2-\sinh^2u+{p_\alpha^2\over\sinh^2u}= a^2\,,\qquad
\theta'^2-\sin^2\theta+{p_\zeta^2\over\sin^2\theta}=-a^2\,.
}
These equations can be solved in general in terms of elliptic
integrals, which we will not do here, since we will use in Section~4.4
a different coordinate system where the solution can be found rather
easily.

Instead, let us focus for now on the simple case with
$p_\alpha=-p_\zeta=0$ which is the case when
the Wilson loop \gaugeWL\ and surface operator \op\ are in phase, {\it i.e.}
$\psi_0=\phi_0$.
One solution is for $a^2=1$ which
corresponds to the Wilson loop on a great circle on
$S^5$.
\eqn\strsol{
\psi-\psi_0=-\phi=\tau\,,\qquad
\cosh u={1\over\cos\sigma}\,,\qquad
\theta={\pi\over2}\,.
}
The coordinate $u$ varies between infinity, the boundary of spacetime, and
$u_0$ where it ends on the D3-brane. After
subtracting the divergence from infinity, the finite part of the string
action is
\eqn\clstac{
\cS={\sqrt{\lambda}\over2}
\int d\sigma(u'^2+\cosh^2u)=-\sqrt\lambda\sinh u_0\,.
}

On the D3-brane worldvolume we have a non-trivial holonomy for the worldvolume
gauge field  $A_\psi=\alpha$ \defkap. Since the string ends on the D3-brane along
a curve wrapping the $\psi$ direction there is a contribution from a boundary term on the worldsheet.
It is given by
\eqn\wilonbrane{
e^{ i{\oint A_\psi\,d\psi}}=e^{2\pi i\alpha}\,.
}

The correlator between a surface operator \op\ and the Wilson loop is given in the semiclassical probe approximation by string worldsheet action. Combining the bulk and boundary term and using the relation \defkap\
between
 $\sinh u_0$ and  the parameters
of the surface operator \op\ in the gauge theory, the expectation
value of the Wilson loop in the presence of the surface operator \op\ is given by
\eqn\wilsurfprob{
{\vev{W_{\pi/2,\phi_0}\cdot \ope}\over\vev{\ope}}
=\exp\left(2\pi|\beta+i\gamma|+2\pi i\alpha\right).
}
This probe D3-brane computation  reproduces   the result 
of the   gauge theory computation \gaugeWLvalue\ for gauge group $U(1)\times SU(N-1)$.

Note that there is another string solution with these boundary
conditions \ZaremboWL, which has $a=0$ and
\eqn\zarembosol{
\sinh u={1\over\sinh\sigma}\,,\qquad
\sin\theta={1\over\cosh\sigma}\,.
}
This solution does not end on the D3-brane, but is still a
solution with the prescribed boundary conditions. This solution
has zero action.

Using the coordinates in the next section and the explicit solution
in Section 4.4, we can evaluate the action for arbitrary
$\psi_0-\phi_0$ and $\theta_0$. There are two disconnected solutions, found
already in \DrukkerCU\ with action
\eqn\disconne{
\cS=\pm\sqrt\lambda\,\cos\theta_0\,.
}
Clearly the solution with a negative sign dominates, but the other one, which 
is unstable, can 
also be found in the asymptotic expansion of the Gaussian matrix model and 
we therefore retain it.

The connected solution, with the contribution of the holonomy on
the probe D3-brane  has action
\eqn\conne{
\cS=-\sqrt\lambda\,\sin\theta_0\cos(\psi_0-\phi_0)\sinh u_0
-2\pi i\alpha\,.
}
Summing over the three saddle points and using \defkap\ we get the
result
\eqn\wilsurfprob{\eqalign{
{\vev{W_{\theta_0,\psi_0}\cdot \ope}\over\vev{\ope}}
=&\exp\big(2\pi \sin\theta_0(\beta\cos\psi_0+\gamma\sin\psi_0)+2\pi i\alpha\big)
+\cr&{}
+\exp\left(\sqrt\lambda\cos\theta_0\right)
+\exp\left(-\sqrt\lambda\cos\theta_0\right).
}}
This agrees precisely with the gauge theory prediction \gaugeWLvalue\ for
$N_1=1$,  $N_2=N-1$ and $\beta_2=\gamma_2=0$ in the large $N$ limit, including 
the last term, corresponding to an unstable solution.

In string theory the difference between a Wilson loop and an 't~Hooft loop
is simply the replacement of a fundamental string by a D-string.
Geometrically the surfaces will look the same with the only difference being
  a relative factor of $1/g_s=4\pi /g^2_{YM}$ in the on-shell action \clstac.
 In this case the D-strings couples through  its boundary  to the dual gauge field on the probe
 D3-brane, which is now proportional    to $\eta$ \defkap.
 Therefore the final answer of the calculation of the correlator
 between a surface operator \op\ and an 't Hooft operator in the
 probe approximation is
\eqn\tHooftsurfprob{\eqalign{
{\vev{T_{\theta_0,\psi_0}\cdot \ope}\over\vev{\ope}}
=&\exp\left({8\pi^2\over g_{YM}^2}\sin\theta_0
(\beta\cos\psi_0+\gamma\sin\psi_0)+2\pi i\alpha\right)
+\cr&
+\exp\left({4\pi N\over\sqrt\lambda}\,\cos\theta_0\right)
+\exp\left(-{4\pi N\over\sqrt\lambda}\,\cos\theta_0\right).
}}
This calculation done with a D1-brane is exactly the $S$-dual
of \wilsurfprob\ accounting also for the $S$ duality transformation
$\alpha\to\eta$.

\newsec{``Bubbling'' Surface Operators in ${\cal N}=4$ SYM}

In this section we perform calculations with surface operators
in $\cN=4$ SYM using the Type IIB supergravity solutions proposed
in \GomisFI\ as the gravitational description of surface operators.
We begin with a summary of the main features of the supergravity
solutions. We then review how to extract correlation functions of
gauge theory operators from an asymptotically \ads\ bulk solution and summarize the main formulas.
  We then calculate using the supergravity solutions the correlators of a surface operator \op\ with
  local operators, with a Wilson and an 't Hooft loop.

\subsec{Surface Operator as ``Bubbling'' Supergravity Solution}

In \GomisFI\ the gravitational description of all the maximally supersymmetric
operators \op\ in ${\cal N}=4$ SYM was found in terms of smooth
ten dimensional solutions of Type IIB supergravity which are
asymptotically $AdS_5\times S^5$.
These supergravity solutions   capture the complete backreaction
of the configuration of D3 branes in  \ads\ considered in the
previous section, which describe a surface operator \op\ in
the probe approximation. Here we summarize the main ingredients
of the solutions needed for the holographic computation of the
correlation functions. See \GomisFI\ for more details about the solution and the identification with the maximally supersymmetric surface operators in ${\cal N}=4$ SYM.

By analyzing the
symmetries of the maximally supersymmetric surface operators in $\cN=4$ SYM one may observe \GomisFI\ that the corresponding ansatz for the dual supergravity solution can be found by performing a double analytic continuation of the LLM ansatz
\LinNB\LinNH\
dual
 to the maximally supersymmetric local operators in $\cN=4$ SYM. The Einstein frame metric
describing a surface operator in ${\cal N}=4$ SYM is given by
\eqn\metric{
ds^2=y\sqrt{2z+1\over 2z-1}ds^2_{AdS_3}+y\sqrt{2z-1\over 2z+1}d\Omega_3+{2y\over \sqrt{4z^2-1}}(d\chi+V)^2+{\sqrt{4z^2-1}\over 2y}(dy^2+dx_idx_i),}
while the RR five-form field strength is given by 
\eqn\fiveform{\eqalign{
F_5=&{1\over 4}\left(d\left[y^2{2z+1 \over 2z-1}(d\chi+V)\right]-y^3\ast _3d\left[{z+{1\over 2}\over y^2}\right] \right)\wedge
d\,V\!ol _{AdS_3}\cr
&-{1\over 4}\left( d\left[y^2{2z-1 \over 2z+1}(d\chi+V)\right]-y^3\ast _3d\left[{z-{1\over 2}\over y^2}\right]\right)\wedge d \Omega _3.}}
$z$ is a function on the space $X$ parametrized by $x_1$, $x_2$ and
$y$, with $ds^2_X=dy^2+dx_idx_i$ and $y\geq0$.
$V$ is a one-form satisfying $dV={1\over y} *_{X}dz$,
and hence, the metric and the
five-form are completely determined by $z(x_1,x_2,y)$.

This supergravity ansatz describes a surface operator \op\ in $\cN=4$
on $AdS_3\times S^1$, as the metric in   the conformal boundary of
\metric\ is that of $AdS_3\times S^1$.
The choice of $AdS_3$ metric determines the surface $\Sigma$ on
which the surface operator  \op\ is supported: $AdS_3$ in Poincar\'e coordinate corresponds to a surface operator \op\ supported on $\Sigma=R^{2}$ while global  $AdS_3$ corresponds to a surface operator \op\ supported on $\Sigma=$S$^2$.

A non-trivial solution to the equations of motion is obtained by
specifying a configuration of $M$ point-like particles in $X$.
The data from which the solution is determined is the particle
positions $(\vec{x}_l,y_l)$ in $X$ (see Figure 1). Given a particle
distribution, the function $z(x_1,x_2,y)$ is found by solving the
following differential equation\foot{The ``charge" $Q_l=2\pi y_l$
associated to these particles is fixed such that the $\chi$ circle shrinks
smoothly at $y=y_l$.}
\eqn\laplace{
\partial_i\partial_i z(x_1,x_2,y)+y\partial_y\left({\partial_y z(x_1,x_2,y)\over y}\right)=\sum_{l=1}^M2\pi y_l\delta(y-y_l)\delta^{(2)}(\vec{x}-\vec{x}_l).}

\ifig\Youngtabcolumnsrows{$a)$ The metric and five-form flux is determined once the position of the particles in $X$
-- labeled by coordinates $(\vec{x}_l,y_l)$ where $y\geq 0$ -- is given. The $l$-th particle is associated with a point $P_l\in X$. $b$) The configuration corresponding to the \ads\ vacuum.}
{\epsfxsize3in\epsfbox{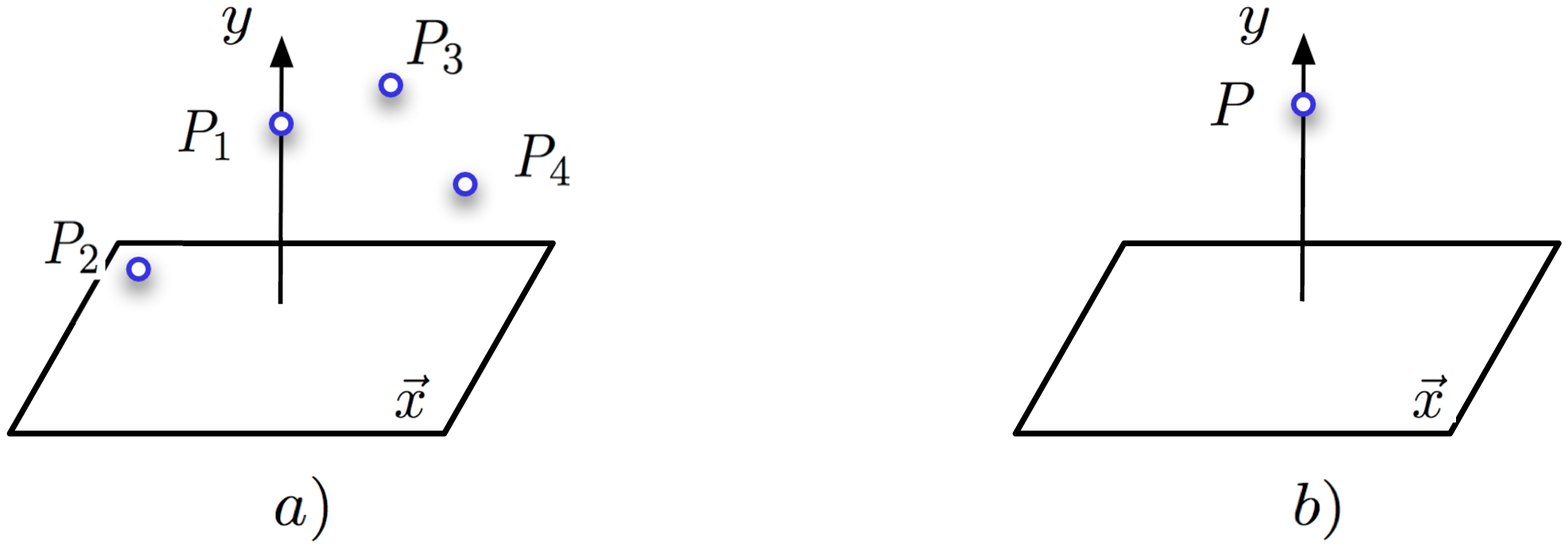}}

This equation is equivalent to the six dimensional Laplace equation with $SO(4)$ invariant sources
for the variable 
\eqn\varia{
\Phi={z\over y^2}.}
The solution to  \laplace\ giving rise to a non-singular smooth metric is given by
\eqn\afin{
z(x_1,x_2,y)={1\over 2}+\sum_{l=1}^M z_l(x_1,x_2,y),
}
where 
\eqn\expansio{
z_l(x_1,x_2,y)={(\vec{x}-\vec{x}_l)^2+y^2+y_l ^2
\over 2
\sqrt{((\vec{x}-\vec{x}_l)^2+y^2+y_l ^2 )^2-4y_l ^2y^2}}-{1\over 2}.}
The one form $V\equiv V_Idx^I$ can be determined up to an exact form by solving $dV={1\over y} *_{X}dz$ 
\eqn\oneform{
V_I=-\epsilon_{IJ} \sum_{l=1}^M
{(x^J-x^J_{l})
((\vec{x}-\vec{x}_l)^2+y^2-y_l ^2 )
\over
2 (\vec{x}-\vec{x}_l)^2 \sqrt{((\vec{x}-\vec{x}_l)^2+y^2+y_l ^2 )^2-4y_l
^2y^2}
}.}
Therefore, the metric and five-form field strength are determined by the integer $M$ and by the particle locations $(\vec{x}_l,y_l)$ for $l=1,\cdots, M$.

As explained in \GomisFI, further data needs to be given in order to fully specify a Type IIB supergravity solution. These metrics have $M$ non-trivial two-cycles $D_l$ and a solution is fully specified only after we fix the periods of the NS-NS and RR two-form gauge fields on the two cycles 
\eqn\holoa{
\int_{D_l} {B_{NS}\over 2\pi};\qquad \qquad\int_{D_l} {B_{R}\over 2\pi};\qquad l=1,\cdots,M.}
\ifig\nontrivialdisks{A disk $D$ can be constructed by fibering S$^1$ over an interval connecting the ``charge" at the point $P_l\in X$ with $(\vec{x}_l,y_l)$ coordinates and the boundary of \ads.}
{\epsfxsize1.5in\epsfbox{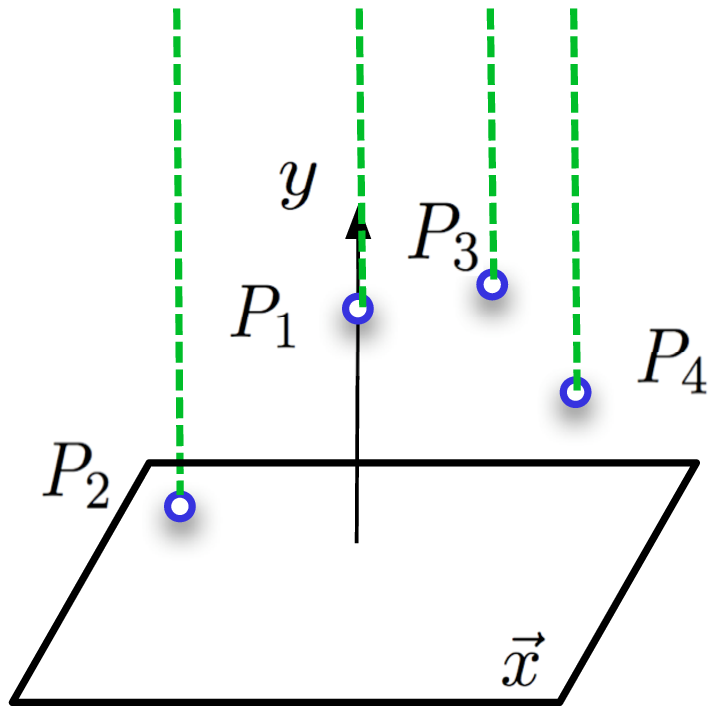}}

Therefore, these ``bubbling'' supergravity solutions depend on $4M$ real parameters and there is a one to one correspondence between these supergravity solutions and the maximally supersymmetric surface operators in ${\cal N}=4$ SYM, which also depend on $4M$ parameters $(\alpha_l,\beta_l,\gamma_l,\eta_l)$.
The explicit mapping is given by \GomisFI 
\eqn\map{\eqalign{
\alpha_l&=-\int_{D_l}{B_{NS}\over 2\pi}\cr
\beta_l+i\gamma_l&=~{x_{l,1}+ix_{l,2}\over 2 \pi l_s^2}\qquad\qquad l=1,\cdots,M\cr
\eta_l&=\int_{D_l}{B_{R}\over 2\pi}\cr
y_l^2&={N_l\over N}.}}
The ranks of the subgroups of the gauge group $L=\prod_{l=1}^MU(N_l)$  left unbroken by \op\  is encoded in the positions in the $y$ axis of the $M$ particles via $N_l={y_l^2\over 4\pi l_p^4}$, as $y_l^2$ determines the amount of five-form flux through the corresponding $S^5$. In the units where the \ads\ radius of curvature $L=1$, we have that $y_l^2=N_l/N$.

These solutions are regular but  become singular precisely when the path integral describing a surface operator \op\ becomes singular  \GomisFI. This  occurs when the unbroken gauge group near $\Sigma$ is enhanced, which corresponds in the supergravity solution to colliding the charges in Figure 1. When two charges collide an $S^2$ in the geometry shrinks to zero size and the solution is singular.

\subsec{Gauge Theory Correlation Functions from Ten Dimensional ``Bubbling'' Solutions}

 Here we summarize the main ideas and formulas involved in extracting the correlation function of local operators in the boundary gauge theory from the ten dimensional ``bubbling'' supergravity solution.
We refer to
\SkenderisYB\ for   more details and derivations.

The asymptotically \ads\ solutions of Type IIB supergravity describing maximally supersymmetric surface operators \op\ in ${\cal N}=4$ SYM are ten dimensional. Holography is on the other hand an equivalence between quantum gravity with $AdS_5$ boundary conditions and a four dimensional field theory on the boundary
\MaldacenaRE\GubserBC\WittenQJ. Therefore,
given a ten dimensional asymptotically \ads\ solution of Type IIB supergravity, one must systematically reduce the ten dimensional solution to a solution of five dimensional gravity coupled to --- in general --- an infinite
number of five dimensional fields. Once this reduction and a local five dimensional action has been found, one can apply the usual holographic rules \MaldacenaRE\GubserBC\WittenQJ\ and use the bulk five dimensional description to compute the correlation function of boundary operators. In order to get the renormalized correlation functions in the gauge theory one must appropriately renormalize the bulk  gravitational action using holographic renormalization
\lref\deHaroXN{
  S.~de Haro, S.~N.~Solodukhin and K.~Skenderis,
  ``Holographic reconstruction of spacetime and renormalization in the
  $AdS$/CFT
  correspondence,''
  Commun.\ Math.\ Phys.\  {\bf 217}, 595 (2001)
  [arXiv:hep-th/0002230].
}
\lref\BianchiDE{
  M.~Bianchi, D.~Z.~Freedman and K.~Skenderis,
  ``How to go with an RG flow,''
  JHEP {\bf 0108}, 041 (2001)
  [arXiv:hep-th/0105276].
}
\lref\SkenderisWP{
  K.~Skenderis,
  ``Lecture notes on holographic renormalization,''
  Class.\ Quant.\ Grav.\  {\bf 19}, 5849 (2002)
  [arXiv:hep-th/0209067].
}
\deHaroXN\BianchiDE\BianchiKW\ (for a review and more references see \SkenderisWP).

The first step in constructing the five dimensional gravity action
from which one can holographically compute gauge theory correlation functions is to decompose  the  ten dimensional solution into  \ads\  plus the deviation of the solution away from \ads. One then expands the deviation in a complete basis of $SO(6)$ spherical harmonics which yields an infinite collection of five dimensional fluctuation fields. The decomposition into harmonics of the   the metric and five-form deviations  is given by
\eqn\kkredun{\eqalign{
h_{\mu\nu}&=\sum h_{\mu\nu}^{I_1}(x)Y^{I_1}(y)\cr
h_{\mu a} (x,y) &=
\sum ({B}^{I_5}_{(v)\mu }(x) Y_a^{I_5}(y)
+ {B}^{I_1}_{(s)\mu}(x) D_a Y^{I_1}(y)) \cr
h_{(ab)}(x,y)
&= \sum (\hat{\phi}_{(t)}^{I_{14}}(x) Y_{(ab)}^{I_{14}}(y)
+ \phi^{I_5}_{(v)}(x) D_{(a} Y^{I_5}_{b)}(y)
+ \phi^{I_1}_{(s)}(x) D_{(a} D_{b)} Y^{I_1}(y) )\cr
h_{a}^a(x,y) &= \sum {\pi}^{I_1}(x) Y^{I_1}(y),
}}
and
\eqn\kkredunda{\eqalign{
f_{abcd\mu}(x,y) &= \sum
(D_\mu b_{(s)}^{I_1}(x) \epsilon_{abcd}{}^e D_e Y^{I_1}(y)
+ (\Lambda^{I_5}-4) b_\mu^{I_5}(x) \epsilon_{abcd}{}^e Y_e^{I_5}(y)) \cr
f_{a b c d e}(x,y) &= \sum b_{(s)}^{I_1}(x) \Lambda^{I_1} \epsilon_{abcde} Y^{I_1}(y),
}}
where $x^\mu$ are $AdS_5$ coordinates while $y^a$ are S$^5$ coordinates.
$Y^{I_1}$, $Y^{I_5} _a$ and $Y_{(ab)}^{I_{14}}$
are scalar, vector and symmetric traceless $SO(6)$ spherical harmonics respectively while
$-\Lambda^{I_1}$ and $-\Lambda^{I_5}$ are the eigenvalues of the
d'Alembertian on
scalar and vector
harmonics (see Appendix~A for more details on the relevant spherical harmonics).
 One must then solve the ten dimensional Type IIB supergravity equations of motion perturbatively
 in the number of fluctuations.

 In \Nieuwen\ the linearized analysis was performed in the de Donder gauge which retains the physical fluctuations. For our solutions it is convenient to work with gauge invariant variables as our  ``bubbling'' solution \metric\    is not in the de Donder gauge. The perturbative analysis using gauge invariant variables has been carried out by Skenderis and Taylor in \SkenderisUY, where they showed that the equations of motion for the gauge invariant variables indeed coincide with the equations of motion in the de Donder gauge if the fluctuations in the de Donder gauge are replaced by their gauge invariant counterparts. The gauge invariant variables to linear order in the fluctuations
are given in terms of the fluctuations in \kkredun\kkredunda\ by \SkenderisYB 
\eqn\gaugeinv{\eqalign{
\hat{\pi}^{I_1} &= {\pi}^{I_1} - \Lambda^{I_1} \phi_{(s)}^{I_1} \cr
\hat{b}^{I_1} &= b^{I_1}_{(s)} - {1\over 2} \phi_{(s)}^{ I_1}\cr
\hat{B}^{I_5}_{(v) \mu} & = {B}^{I_5}_{(v)\mu}- {1\over 2} D_\mu \phi^{I_5}_{(v)}\cr
\hat{b}{}^{I_5}_{\mu} &= b_{\mu}^{I_5} - {1\over 2(\Lambda^{I_5}-4)} D_\mu \phi_{(v)}^{I_5}\cr
\hat{\phi}^{I_{14}}&={\phi}^{I_{14}}.}}
The gauge invariant fluctuations that diagonalize the equations
of motion to leading order in fluctuations and that couple to
the CPO's $\cO_\Delta^I$ with  $\Delta=2,3$, the $SO(6)$ $R$-symmetry current $J^a_m$ and the stress-energy tensor $T_{mn}$ of $\cN=4$ SYM are \WittenQJ  
\eqn\diagonfluct{\eqalign{
\hat{s}^{2I}&={1\over 80}\left(\hat{\pi}^{2 I} -60 \hat{b}^{2 I}\right)\cr
\hat{s}^{3I}&={1\over 100}\left(\hat{\pi}^{3 I} -70 \hat{b}^{3 I}\right)\cr
A_{\mu}^a&=\hat{B}^{0}_{(v) \mu}-12 \hat{b}{}^{0}_{\mu}\cr
\tilde{h}_{\mu\nu}&={h}_{\mu\nu}^0+{1\over 3}g_{\mu\nu}^0\pi^0.}}

In order to calculate correlation functions in the boundary gauge
theory from the bulk description one must find a local five
dimensional action that controls the dynamics of the gauge
invariant fluctuations dual to the gauge theory operators.
This requires finding a non-linear map between local five
dimension fields and the gauge invariant fluctuations obtained
by expanding the ten dimensional equations of motion. This
mapping is constructed such  that a solution of the five
dimensional equations of motion yields a solution of the ten
dimensional Type IIB supergravity equations of motion.
This mapping can be
constructed perturbatively in the number of fluctuations.
This non-linear map between local fields and gauge invariant
fluctuations has been constructed in \SkenderisUY.

Once a local five dimensional action yielding the appropriate five
dimensional equations of motion is constructed, the correlation
functions of renormalized boundary operators can be computed
using holographic renormalization. In order to construct the
renormalized correlation function of boundary operators the
bulk action must be enriched with a set of boundary terms to
make the on-shell bulk action finite. Despite the fact that
the five dimensional action in general contains an infinite
number of fields, only a finite number of terms in the action
are infrared divergent. It is only the bulk  fields dual to
low dimension operators in the gauge theory that require
boundary counterterms. Once a complete set of counterterms
for the five dimensional action are constructed, one can calculate the correlation functions of boundary operators by taking functional derivatives of the renormalized bulk gravitational action.

Here we are interested in computing the correlation  function of
the CPO's ${\cal O}^I_\Delta$ with $\Delta=2,3$, of the $R$-symmetry  current $J_m^a$ and of the stress-energy tensor $T_{mn}$   with a surface operator \op\ in ${\cal N}=4$ SYM. The bulk action that captures the degrees of freedom dual to these operators is given by \BianchiDE \SkenderisUY\foot{This action is written, as throughout the rest of the paper, in units where the radius of curvature of \ads\ is $L^4=4\pi g_sNl_s^4=1$.}
\eqn\bulkaction{\eqalign{
S_{bulk}={N^2\over 2\pi^2}\int\ d^5x \sqrt{-G}\,\Bigg(& {R\over 4} +{1\over 2}G^{\mu\nu}\partial_{\mu}\Theta \partial_{\nu}\Theta+{1\over 2}G^{\mu\nu}\partial_{\mu}\Psi \partial_{\nu}\Psi-2\Theta^2-{3\over 2} \Psi^2\cr
&-{4\sqrt{6}\over 3}\Theta^3-{1\over 16} F^a_{\mu\nu}F^a_{\mu\nu}\Bigg),}}
where $F_{\mu\nu}^a$ is the field-strength of a vector potential
$A_\mu^a$ and
$G_{\mu\nu}$, $\Theta$, $\Psi$ and $A^a_{\mu}$ couple to
$T_{mn}$, ${\cal O}^I_{\Delta=2}$, ${\cal O}_{\Delta=3}^I$ and $J^a_m$
respectively and $a$ is an $SO(6)$ vector index.
The boundary counterterms required to have a finite on-shell action
and from which the correlation function of renormalized operators
can be computed can be found by solving the equations of motion
of \bulkaction\ near the boundary of $AdS_5$. They are written down
in \BianchiDE \SkenderisUY.

In order to calculate the one point function of these local operators
using the bulk description it is convenient to write the five
dimensional metric $G_{\mu\nu}$ in the Fefferman-Graham form
\eqn\fg{
ds_5^2={dU^2\over U^2}+{1\over U^2}\left( G_{(0){mn}}+U^2G_{(2){mn}}+\cdots\right)dx^mdx^n,}
where in our case $G_{(0)}$ is the $AdS_3\times S^1$ metric on the conformal boundary at $U=0$, where $\cN=4$ SYM lives.
The bulk fields $\Theta, \Psi$ and $A_{\mu}$  also have a near boundary expansion. It is given by\foot{In our solutions the bulk fields do not have a source term and therefore we do not write it. In a general solution a non-normalizable mode has to be added to these asymptotic solutions.}
\eqn\nearbound{\eqalign{
\Theta\equiv \hat{s}^{2I}(x,U)&=U^2\left[\hat{s}^{2I}\right]_2Y_2^I+\ldots\cr
\Psi\equiv \hat{s}^{3I}(x,U)&=U^3\left[\hat{s}^{3I}\right]_3Y_3^I+\ldots\cr
A^a_{\mu}(x,U)&=U^2\left[A_{\mu}^{a}\right]_2(x)Y^{1a}+\ldots,}}
where $\left[ \,\cdot\, \right]_{.}$ is the normalizable mode of the field.

Evaluation of the on-shell one point function of the renormalized bulk action shows that the vacuum expectation value of these operators is  captured by the normalizable mode of the dual bulk field.
In order to calculate the one point function of a unit normalized operator, we must divide the one point function obtained from supergravity  by the square root of the two-point function obtained from supergravity \LeeBXA, such that the   supergravity computation of the vacuum expectation value corresponds to that of a  unit normalized operator with two-point function given by \twopoint.
 After taking this into account, we have that the one
 point function of the CPO's $\cO^I_\Delta$ with $\Delta=2,3$ and of the $R$-symmetry current are given by \BianchiDE \SkenderisUY\foot{The current $J^a _{\mu}$ is normalized so that it has the canonical two-point function \BianchiDE.}
\eqn\chiralvec{\eqalign{
\vev{{\cal O}_{2}^I}_{\ope}=&{N \over2 }{2\sqrt{8}\over3}
\left[\hat{s}^{2I}\right]_2,\cr
\vev{{\cal O}_{3}^I}_{\ope}=&{3N \over2\sqrt{6} }
\left[\hat{s}^{3I}\right]_3,\cr
\vev{J^a _{\mu}}_{\ope} =&-{\sqrt{2}N^2\over 24\pi^2}\left[ A_{\mu}^{a}\right]_2.}}

The computation of the vacuum expectation value of the stress-energy tensor $T_{mn}$ is more subtle as  the bulk scalar field $\Theta(x,U)$ enters in the renormalization procedure for the five dimensional metric. The formula for the vacuum expectation value of the stress-energy tensor is given by \BianchiDE \SkenderisUY 
\eqn\stress{\eqalign{
\left< T_{mn} \right>_{\ope} &= {N^2\over2 \pi^2}\left(
g_{(4)mn} - {2\over9}  \left[\hat{s}^{2I}\right]_2\left[\hat{s}^{2I}\right]_2  g_{(0)mn} \right . \cr
& \left . + {1\over8}[\Tr g_{(2)}^2 -(\Tr g_{(2)})^2] g_{(0)mn}
- {1\over2} (g_{(2)}^2)_{mn} + {1\over4} g_{(2)mn} \Tr g_{(2)}
+{3\over2} h_{(4)mn} \right)
.}}
In this formula $g_{(\cdot)mn}$ denotes the various terms in the near boundary expansion of the ten dimensional metric in the Graham-Fefferman coordinate $z$ in \fg.

As explained in Section $2$, the expectation value of a local operator $\vev{\cO}_{\ope}$ in a non-trivial state of the gauge theory on $AdS_3\times S^1$ corresponds to the two-point correlation function of the surface operator \op\ with the local operator $\cO$ in $\cN=4$ SYM on $R^4$ 
\eqn\opeeee{
{\vev{\ope\cdot{\cal O}}\over \vev{\ope}}.}
Therefore, we can calculate these correlators computed in the previous sections in other regimes by using the supergravity description of surface operators.

\subsec{Correlator of Surface Operators with Local Operators from ``Bubbling'' Solutions}

We can now compute the correlation function of a surface operator
\op\ with the CPO's $\cO^I_\Delta$ of $\Delta=2,3$, with the $U(1)$ current $J_m^\psi$ and with  the stress-energy tensor $T_{mn}$ of ${\cal N}=4$ SYM using the holographic correspondence. The analogous calculation for local operators has been carried out by Skenderis and Taylor in
\SkenderisYB. As summarized in the previous subsection,
the computation of the correlator of \op\  with these local operators
requires extracting the appropriate coefficient of the near boundary expansion of the corresponding fluctuation mode in \chiralvec\stress\ from the ``bubbling'' supergravity solution  \metric\fiveform.

Before looking at the fluctuations, we write down a reference
\ads\ solution, which is the metric \metric\fiveform\ with a single
particle source. If the particle position is at
$({\vec x}^{(0)},y^{(0)}=1)$, then in the coordinates
\eqn\rela{\eqalign{
x^1-x^{(0)}_1&+i(x^2-x^{(0)}_2)=r e^{i\alpha}\cr
r&=  R\sin\theta\cr
y&= \sqrt{1+R^2} \cos\theta\cr
\chi&={1\over 2}(\psi-\phi)\cr
\alpha&=\psi+\phi,}}
we have that:
\eqn\backrG{\eqalign{
z^{(0)}&={1\over 2}{R^2+\cos^2\theta+1\over R^2+\sin^2\theta}\cr
V^{(0)}&={1\over 2}{R^2-\sin^2\theta\over R^2+\sin^2\theta}\,d\alpha.}}
Inserting \backrG\ into \metric\ results in the \ads\ metric with
$AdS_5$ foliated by $AdS_3\times$S$^1$ slices\foot{
This is the same as \metsphereII\ with $R=\sinh u$.}
\eqn\metricads{
ds^2= \left[ (1+R^2) ds^2_{AdS_3}+{dR^2\over 1+R^2}+R^2 d\psi^2+\cos^2\theta d\Omega_3+d\theta^2+\sin^2\theta d\phi^2\right].} The five-form flux \fiveform\ is given by 
\eqn\fluxvacc{
F_5=R(R^2+1)dR\wedge d\psi\wedge d\,V\!ol_{AdS_3}+\cos^3\theta \sin\theta  d\phi \wedge d\theta\wedge d\Omega_3.}

The supergravity solution \metric\fiveform\ describing a maximally supersymmetric
surface operator \op\ is characterized by a choice of particle
distribution $(\vec x_l,y_l)$ with $l=1,\cdots,M$. This
determines the function $z(x_1,x_2,y)$ and the one form $V$.
For computational purposes we find it useful to use the function
$\Phi$ \varia, which is given in terms of $z(x_1,x_2,y)$ by 
\eqn\variade{
\Phi={z\over y^2}.}
We    then rewrite $\Phi(x_1,x_2,y)$ and $V_I(x_1,x_2,y)$ as
deviations from the reference \ads\ solution
\eqn\nearads{\eqalign{
\Phi&=\Phi^{(0)}+\Delta \Phi\,,\qquad
\Phi^{(0)}\equiv {z^{(0)}\over y^2}\,,\cr
V&=V^{(0)}+\Delta V.}}
It turns out that the most useful reference metric, around
which the fluctuations are minimal, is the one centered around
the center of mass defined by\foot{
Recall \map\ that $y_l^2=N_l/N$, and is therefore a natural
weight for the point $\vec x_l$.}
\eqn\cm{
\vec x^{(0)}=\sum_{l=1}^M y_l^2{\vec x}_l\,,\qquad
\sum_{l=1}^M y_l^2=1.}
Furthermore, to simplify the calculations, henceforth we  take
that $\vec x^{(0)}=0$.

Now, since the  function $\Delta\Phi$ solves the six dimensional
Laplace equation with $SO(4)$ symmetry, all the fluctuations can
be expanded in a basis of  $SO(4)$  invariant harmonics
$Y^{\Delta,k}$ of $SO(6)$ (see Appendix~A   for  a summary of
their properties). Therefore
\eqn\DeltaPhiexp{
\Delta\Phi=\sum_{\Delta\geq 2}\
\sum_{k=-\Delta,-\Delta+2,\cdots}^\Delta\Delta\Phi_{\Delta,k}\,
{Y^{\Delta,-k}(\theta,\phi) \over R^{\Delta+4}}
,}
where $\theta,\phi$ are coordinates on the $S^5$ \metricads.
The $\Delta=1$ term in \DeltaPhiexp\ vanishes in the center
of mass frame which we have adopted.

By doing the calculation in the center of mass frame \cm\ with
$\vec x^{(0)}=0$,
we get from the $M$ particle solution \afin\expansio\ that%
\foot{
These expression will also be valid away from the center
of mass frame, with a subtlety appearing only in $\Delta\Phi_{3,\pm1}$, where 
the factor of $1/2$ needed to be determined by a separate calculation.}
\eqn\DeltaPhimoments{\eqalign{
\Delta \Phi_{2,0}&=
4\sqrt{3}\sum_{l=1}^M\left(\vec{x}^2_l-{y_l^2-1\over2}\right)y^2_l\cr
\Delta \Phi_{2,\pm2}&=
6e^{\mp 2i\psi} {\sum_{l=1}^M(x_{l,1}\pm i x_{l,2})^2y^2_l}\cr
\Delta \Phi_{3,\pm1}&=
8\sqrt{3}e^{\mp i\psi}
\sum_{l=1}^M(x_{l,1}\pm i x_{l,2})(\vec{x}_l^2-y_l^2+{1\over 2})y_l^2
\cr
\Delta \Phi_{3,\pm3}&=8\sqrt{2}e^{\mp i3\psi}
\sum_{l=1}^M(x_{l,1}\pm i x_{l,2})^3y_l^2.
}}

Now to calculate the desired expectation values of $\cO_{\Delta,k}, J_m^a$ and $T_{mn}$
we   use the  relation \chiralvec\stress\ between the expectation values  and
the fluctuations $\left[\hat{s}^{2I}\right]_2, \left[\hat{s}^{3I}\right]_3,\left[A_\mu^a\right]_2$ and $g_{(\cdot)mn}$. These in turn are related to the
moments of $\Delta\Phi$ which we have just calculated
by following the procedure in Section~4.2 as we outline now.

We need to find the asymptotic form of the metric and five-form
and then expand it in spherical harmonics as in \kkredun\kkredunda.
This is done by plugging the coordinate transformation \rela\ into
the $M$ particle solution \afin\expansio\oneform\ and expanding
for large $R$. In the multi-particle case this will not yield
a manifestly asymptotically \ads\ metric. This can be fixed by
modifying the one-form $V$ \oneform. Recall that
$V$ is determined from $z$ by solving the
equation $ydV=*_X dz$ and hence is only determined up
to an exact form. In order to make manifest the \ads\ asymptotics
we add to $V$ in \oneform\ an exact one form 
\eqn\shift{
V\rightarrow V+d\omega.}
The explicit expression for $\omega$ that yields the desired
asymptotics is given in Appendix~D.

Now we can compute the various fluctuation modes. The warp
factors in the metric \metric\ can be parametrized as follows 
\eqn\dofabcd{\eqalign{
y\sqrt{{2z-1\over2z+1}}
&\equiv
\cos^2\theta(1+A),
\qquad\qquad\qquad
y\sqrt{{2z+1\over2z-1}}
\equiv
(R^2+1)(1+B),\cr
{2y\over \sqrt{4z^2-1}}
&\equiv
(R^2+\sin^2\theta)(1+C),
\qquad\quad\
{\sqrt{4z^2-1}\over 2y}
\equiv{(1+D)\over (R^2+\sin^2\theta)}
.}}
The functions $A, B, C, D$ encode the deviations of the metric away
from $AdS_5\times$S$^5$. We express them in terms of $\Delta\Phi$,
which scales (in the center of mass frame) like $1/R^6$ for
large $R$ \DeltaPhiexp. Up to ${\cal O}(1/R^4)$ they are given by 
\eqn\exofabcd{\eqalign{
 &A=
{1\over2}R^4\Delta\Phi+R^2\sin^2\theta\Delta\Phi-{1\over 8}R^8(\Delta \Phi)^2 \cr
 &B=
 -{1\over2}R^4\Delta\Phi-R^2\sin^2\theta\Delta\Phi+{3\over 8}R^8(\Delta \Phi)^2 \cr
 &C=
-{1\over2}R^4\Delta\Phi-R^2\Delta\Phi+{3\over 8}R^8(\Delta \Phi)^2 \cr
 &D=
{1\over2}R^4\Delta\Phi+R^2\Delta\Phi-{1\over 8}R^8(\Delta \Phi)^2.
}}
We also expand the five-form flux \fiveform\   in a large $R$ expansion. The expression for $\Delta\Phi$ is obtained from the explicit expression for  the $M$ particle solution \afin\expansio.

We now need to use \kkredun\kkredunda\ to calculate the fluctuation
modes that appear in \chiralvec\stress\ which are needed for the
computation of our  correlation functions.
Recalling the expression for the gauge invariant fluctuations
\diagonfluct, we finally have that 
\eqn\sformulas{\eqalign{
\left[\hat{s}^{2,k}\right]_2
&={1\over 8}\left(\Delta\Phi \right)_{2,k}\,,\qquad\quad k=2,0,-2\cr
\left[\hat{s}^{3,k}\right]_3
&={1\over 12}\left(\Delta\Phi \right)_{3,k}\,,\qquad k=3,1,-1,-3\cr
\left[\hat{A}_{\mu}^{\phi}\right]_1&=-{3\over \sqrt{6}}\left(\Delta\Phi \right)_{2,0}.}}

Now we recall the expressions for the moments of $\Delta\Phi$,
which we calculated in \DeltaPhimoments\ and can immediately
write down the supergravity answer for the
correlator of a surface operator \op\ with the CPO's $\cO_{\Delta,k}$
\opers\ with $\Delta=2,3$. Writing the answer in  gauge theory variables using \map\  we get that 
\eqn\finalbubbles{\eqalign{
{\vev{\ope\cdot {\cal O}_{2,0}}\over \vev{\ope}}&={1\over |z|^2}
{8\pi^2\over\sqrt{6}  \lambda}
\left(
\sum_{l=1}^M N_l\left((\beta_l^2+\gamma_l^2)
+{\lambda\over 4\pi^2}{N-N_l\over 2N}\right)\right),
\cr
{\vev{\ope\cdot {\cal O}_{2,2}}\over \vev{\ope}}&={1\over z^2}
{4\pi^2\over \sqrt{2}\lambda}\sum_{l=1}^M
N_l(\beta_l+ i\gamma_l)^2,
\cr
{\vev{\ope\cdot {\cal O}_{3,1}}\over \vev{\ope}}&={1\over z|z|^2}
{8\pi^3\over\sqrt{2}  \lambda^{3/2}}
\left(
\sum_{l=1}^M N_l\left({(\beta_l^2+\gamma_l^2)}
+{\lambda\over 4\pi^2} {N-2N_l\over 2N}\right)(\beta_l+ i\gamma_l)\right),
\cr
{\vev{\ope\cdot {\cal O}_{3,3}}\over \vev{\ope}}&={1\over z^3}
{8\pi^3\over \sqrt{3}\lambda^{3/2}}\sum_{l=1}^M
N_l(\beta_l+ i\gamma_l)^3.}}
The remaining correlators can be obtained by complex conjugating \finalbubbles. 

To compare with the probe brane calculation \probcpofin\ in the
previous section, we consider the result
of the supergravity calculation \finalbubbles\ for  $N_1=1$ and
$N_2=N-1$. Then we impose the center of mass condition, which
is valid for the probe calculation and finally take
$N\rightarrow \infty$ limit. The results are in precise agreement.

We can compare with the gauge theory for arbitrary surface operator parameters 
$(N_l,\,\beta_l,\,\gamma_l)$. As in the case of the probe calculation 
\probcpofin, the supergravity computation matches exactly the gauge theory result 
\correlafin\   for operators with $\Delta=|k|$ and the semiclassical gauge theory result in 
\correlafin\ exactly matches the leading term in the supergravity computation \finalbubbles\ for the rest of the chiral primary operators. 
The supergravity computation suggest that the loop corrections to the gauge theory result  for the correlator between \op\ and $\cO_{\Delta,k}$ truncate  at order $\lambda^{(\Delta-|k|)/2}$.

The vacuum expectation value of the $U(1)$ current  $J_{m}^\psi$ in given by\foot{In computing the correlator with $J_m^\psi$ we have used that the supergravity solution is $U(1)$ invariant and that this $U(1)$ is the diagonal sum of the $U(1)\subset SO(1,5)$ generated by $J_m^\psi$ which rotates the space transverse to $\Sigma$  and the $U(1)\subset SO(6)$ subgroup of the
$R$-symmetry generated by $J_m^\psi$. The $U(1)$ symmetry implies that $\vev{J_m^\psi}_{\ope}=\vev{J_m^\phi}_{\ope}$.}
\eqn\vevcurrent{
\vev{J_\psi^\psi}_{\ope} ={2N\over \lambda}
\left(
\sum_{l=1}^M N_l{(\beta_l^2+\gamma_l^2)}
+{\lambda\over 4\pi^2}{N-N_l\over 2N}\right).}
The first term agrees precisely with the gauge theory result \opecurre\ and the second captures the one loop correction.

We now proceed to calculate the correlator of \op\ with the stress-energy tensor $T_{mn}$. We first calculate the near boundary expansion of the metric along the $AdS_5$ directions. It is given by
\eqn\metricaaa{\eqalign{
ds^2=&d\psi^2\left( R^2+{\Delta\Phi_{2,0}\over 4\sqrt{3}R^2}+{(\Delta \Phi_{2,k}\Delta \Phi_{2,-k})\over 6\times 24 R^2} \right)+{dR^2\over R^2+1}
\left( 1-{\left(\Delta\Phi_{2,k} \Delta\Phi_{2,-k} \right)\over 3\times24 R^4} \right)
\cr
&+ds^2_{AdS_3}\left(R^2+1-{\Delta \Phi_{2,0}\over 12\sqrt{3}R^2}
+{(\Delta \Phi_{2,k}\Delta \Phi_{2,-k})\over 6\times24R^2}\right),
}}
where we sum over the $U(1)$ charge $k=2,0,-2$.

The correlator of \op\ with $T_{mn}$ can be obtained from formula \stress. This requires bringing the ten dimensional metric into the Fefferman-Graham form \fg. In order to bring the metric \metricaaa\ into this form we must perform the following change of coordinates 
\eqn\coordchange{\eqalign{
U=&{1\over R}\left(1-{1\over4R^2}+{1\over R^4}\left(
{1\over 8}-{(\Delta \Phi_{2,k}\Delta\Phi_{2,-k})\over24\times 24}
\right) \right)\cr
R=&{1\over U}\left(1-{1\over 4}U^2-
{(\Delta \Phi_{2,k}\Delta\Phi_{2,-k})\over24\times 24}U^4
\right).
}}
In the Fefferman-Graham gauge the metric \metricaaa\ is given by 
\eqn\metricfin{\eqalign{
ds^2&=d\psi^2
\left({1\over U^2}-{1\over2}+{18+24\sqrt{3}\Delta\Phi_{2,0}+\left(\Delta\Phi_{2,k} \Delta\Phi_{2,-k} \right)
\over 288}U^2 \right)+
{dU^2\over U^2}\cr
&+
 ds^2_{AdS_3}\left({1\over U^2}+{1\over2}+{18-
8\sqrt{3}\Delta \Phi_{2,0}+(\Delta \Phi_{2,k}\Delta \Phi_{2,-k})\over 288}U^2\right).
}}

Now we can read off from \metricfin\ the various terms in the metric
($g_{(4)}$, $g_{(2)}$, $g_{(0)}$) that enter in the formula of the correlator of \op\ with the stress-energy tensor $T_{mn}$ \stress. The correlator is given by
\eqn\stressss{\eqalign{
\vev{T_{\psi\psi}}_{\ope}=&{N^2\over2\pi^2}\left(-{3\over 16}+{1\over4\sqrt{3}}
\Delta\Phi_{2,0}\right),\cr
\vev{T_{ab}}_{\ope}=&{N^2\over2\pi^2}\left({1\over 16}-{1\over12\sqrt{3}}
\Delta\Phi_{2,0}\right)g_{ab},}}
where $g_{ab}$ is an $AdS_3$ metric and the explicit expression for $\Delta\Phi_{2,0}$ is given in \DeltaPhimoments.
More explicitly
\eqn\dimenfinall{\eqalign{
\vev{T_{\psi\psi}}_{\ope}&=-{3N^2\over 32\pi^2} +{2\over g^2_{YM}}{\sum_{l=1}^MN_l\left((\beta_l^2+\gamma_l^2)+{\lambda\over 4\pi^2} {N-N_l\over 2N}\right)}\cr
\vev{T_{ab}}_{\ope}&=\left({N^2\over 32\pi^2}-{2\over  3g^2_{YM}}{\sum_{l=1}^M N_l\left((\beta_l^2+\gamma_l^2)+{\lambda\over 4\pi^2} {N-N_l\over 2N}\right)}\right)g_{ab}\,.}}
The $\lambda$ independent term encodes the Casimir energy in the gauge theory.
The leading term in $\lambda$ precisely agrees with the gauge theory calculation in the semiclassical regime \dimen. The last term can be interpreted as the one loop correction to the gauge theory result \dimen, and suggests that the correlator of \op\ with the stress-energy tensor $T_{mn}$ receives only a one loop correction.

We note that the trace is $\vev{T_m^m}_{\ope}=0$, which  reproduces the fact that
 the conformal anomaly of $\cN=4$ SYM on $AdS_3\times S^1$ vanishes.

\subsec{Correlator with Wilson and 't Hooft Loop}

We turn now to calculating the Wilson and 't Hooft loop expectation
value in the supergravity background  describing a  surface
operator \op. As usual in the holographic
dual, the Wilson loop is described by a classical string
\ReyIK\MaldacenaIM. As in the calculation in the probe approximation
in Section~3.4, we have to identify the boundary conditions on the
string stemming from the parameters of the Wilson loop $W_{\theta_0,\psi_0}$
\gaugeWL\ and 't Hooft loop $T_{\theta_0,\psi_0}$. Unlike the probe approximation, here the string
has no D3-branes to end on, rather it will wrap a non-trivial cycle
in spacetime and close smoothly on itself, as we explain shortly.

To understand the boundary conditions we look at the asymptotic
form of the metric \rela\metricads\ and impose the same conditions
as in the probe calculation in Section~3.4. At the boundary,
where $R\to\infty$, the string should wrap a circle of
$S^5$ at $\theta=\theta_0$, so the coordinates $r$ and $y$
in \rela\ should diverge with a fixed ratio
$r/y\to\tan\theta_0$.
The string wraps circles both in $AdS_5$ and on $S^5$, parametrized
by $\psi$ and $\phi$ with a relative phase $\alpha=\psi+\phi=\psi_0$, corresponding to that of
the Wilson loop $W_{\theta_0,\psi_0}$  \gaugeWL.

To get a smooth string solution with these boundary
conditions, it has to have the topology of the disc and close
smoothly on itself. Recall that the $\chi$ circle shrinks to
zero size at each of the sources $(\vec x_l,y_l)$, so the
string should extend from the boundary to one of the sources.

We take the ansatz where only $\chi$ depends on $\tau$
\eqn\metstringanz{
x_i=x_i(\sigma)\,,\qquad
y=y(\sigma)\,,\qquad
\chi=\tau+\eta(\sigma)\,.
}
Using the metric \metric, the string action in the conformal gauge is
\eqn\metstlagconf{
\cS={\sqrt\lambda\over4\pi}\int d\sigma\,d\tau
\left({2y\over \sqrt{4z^2-1}}
\left((\eta'+V_\sigma)^2+1\right)
+{\sqrt{4z^2-1}\over 2y}(y'^2+x'_ix'_i)\right),
}
where $V_\sigma$ is the pullback of the one-form $V$ in the $\sigma$
direction.

There are two Virasoro constraints for the diagonal and off-diagonal
pieces in the stress-energy tensor
\eqn\metvira{\eqalign{
\eta'+V_\sigma&=0\cr
y'^2+x'_ix'_i&={4y^2\over {4z^2-1}}\,.
}}
The first equation will give $\eta$, once we solve for $V$ (which
depends only on $(\vec x,\,y)$). Then plugging the second Virasoro
constraint into the equations of motion for the $x_i$ and $y$ one finds
\eqn\threeconst{
\left({\sqrt{4z^2-1}\over 2y}\,x_1'\right)'=
\left({\sqrt{4z^2-1}\over 2y}\,x_2'\right)'=
\left({\sqrt{4z^2-1}\over 2y}\,y'\right)'=0\,.
}
These equations mean that the ratios of $y'$, $x_1'$ and $x_2'$ are
constant along the worldsheet. Therefore the string is given
by a straight line in this space!

These equations are actually unnecessarily complicated, since they
are in the conformal gauge. We can switch to the Nambu-Goto
formulation, the the string action is very simple
\eqn\metstlagNG{
\cS={\sqrt\lambda\over2\pi}\int d\sigma\,d\tau
\sqrt{y'^2+x'_ix'_i},
}
In this formulation $\eta$ does not appear and can be set to zero.
The action is clearly just the length functional in
$(\vec x,\,y)$ space, which is solved by straight lines. These worldsheets 
correspond to the disks spanned by the straight lines in Figure 2.

As mentioned above, the solution will have to extend from the boundary
of spacetime to one of the sources, where the $\chi$ circle shrinks away.
The length of the ray emanating from the source $(\vec x_l,\,y_l)$
in the direction of the boundary point
\eqn\bndrypt{
(R\sin\theta_0\cos\psi_0,\,R\sin\theta_0\sin\psi_0,\,
R\cos\theta_0)\,,
}
with $R\to\infty$
is given by
\eqn\divdisty{
R-y_l\cos\theta_0-\sin\theta_0(x_{l,1}\cos\psi_0+x_{l,2}\sin\psi_0)\,.
}

Recall that the ``bubbling" supergravity   background  also has an NS-NS two-form gauge field so in addition
to the Nambu-Goto term, the action includes the flux through the
worldsheet, which according to \map, for a string ending on the
source $l$ is $-2\pi\alpha_l$.

After integrating over $\tau$ and removing the usual infrared divergence,
which here is proportional to the cutoff $R$, the total action is
\eqn\divdist{
\cS
=-\left(y_l\cos\theta_0
+\sin\theta_0(x_{l,1}\cos\psi_0+x_{l,2}\sin\psi_0)
+2\pi i\alpha_l\right).
}
Finally we use \map\ to rewrite the result in terms of the
parameters from the gauge theory description, and
sum over the different classical solutions, one for
each source to get
\eqn\metstringmatch{
{\vev{\ope\cdot W_{\theta_0,\psi_0}}\over\vev{\ope}}\simeq
\sum_{l=1}^M
\exp\left[\sqrt{\lambda N_l\over N}\cos\theta_0
+2\pi \sin\theta_0(\beta_l\cos\psi_0+\gamma\sin\psi_0)
+2\pi i\alpha_l\right].
}

As was mentioned in Section~2.4, when expanding the exact result of the matrix 
model  that captures the supersymmetric Wilson loops one finds two saddle points, 
one with a positive exponent of $\sqrt\lambda$ and one with a negative one. 
In \DrukkerGA\ this was reproduced from a second string solution, an unstable 
one. Such solutions exist also in our case. The reason is that the $(\vec x,\,y)$ 
space is restricted to $y\geq0$, but the plane at $y=0$ is not a boundary. 
In fact, if we consider a string with the ansatz \metstringanz, it would 
simply be reflected from the $y=0$ plane.

We therefore have to amend \divdisty. There isn't a unique geodesic from 
the boundary point \bndrypt\ to each of the sources $(\vec x_l,\,y_l)$, but two. 
The second one is reflected through the $y=0$ plane and its divergent length is%
\foot{This is immediate to derive by using the method of images with a 
source at the point $(\vec x_l,\,-y_l)$.}
\eqn\divdistylong{
R+y_l\cos\theta_0-\sin\theta_0(x_{l,1}\cos\psi_0+x_{l,2}\sin\psi_0)\,.
}
In addition to \metstringmatch\ we therefore find another sum coming from 
the other saddle points, which should all be unstable
\eqn\metunstable{
{\vev{\ope\cdot W_{\theta_0,\psi_0}}\over\vev{\ope}}\simeq
\sum_{l=1}^M
\exp\left[-\sqrt{\lambda N_l\over N}\cos\theta_0
+2\pi \sin\theta_0(\beta_l\cos\psi_0+\gamma\sin\psi_0)
+2\pi i\alpha_l\right].
}
Note that the sign change is only on the first term in the exponent.

The sum of \metstringmatch\ and \metunstable\ 
is in precise agreement with the result of the gauge
theory calculation \gaugeWLvalue\ including our expectation for the term coming
from self-contractions of the extra scalar $\phi^1$. We could
not reproduce the prefactors which
would require a string one-loop calculation to determine.

For an 't~Hooft loop $T_{\theta_0,\psi_0}$ we get the same action, 
except that the string tension is
multiplied by a factor of $1/g_s=4\pi/g_{YM}^2$ and instead of coupling
to the NS-NS field it gets a term from the RR field. The correlator is
\eqn\metthooftmatch{
{\vev{\ope\cdot T_{\theta_0,\psi_0}}\over\vev{\ope}}\simeq
\sum_{l=1}^M
\exp\left[4\pi\sqrt{NN_l\over\lambda}\cos\theta_0
+{8\pi^2\over g_{YM}^2}(\beta_l\cos\psi_0+\gamma\sin\psi_0)\sin\theta_0
+2\pi i\eta_l\right],
}
to which again we should add the contributions of the unstable saddle points. 
This result is $S$-dual to \metstringmatch, accounting for $\alpha_l\to\eta_l$ and 
also matches the general expectations for the behavior of the 't Hooft loop as 
discussed in Section~2.

\newsec{Discussion}

In this paper we have began to study the properties of surface operators \op\ in 
$\cN=4$ SYM performing various explicit calculations with them. We have studied the 
properties of surface operators \op\ using three different realizations, consisting 
of the gauge theory description, the probe D3-brane description and the ``bubbling" 
supergravity description. We have found that a rather nice story emerges between 
these different realizations, allowing for precise matching between quantities 
computed using different descriptions. 

For the expectation value of the surface operator with $\Sigma=R^2$ we found from 
the gauge theory and from the probe approximation that the expectation value is unity.
It would be interesting to learn how to perform this calculation in the framework of 
the ``bubbling'' solutions, but it would probably be more advantageous to do it in the 
case of the surface operators in the  six dimensional theory dual to M-theory on 
$AdS_7\times S^4$ \LuninAB, since unlike our case, those operators do suffer from conformal anomalies.

For the correlator of a surface operator \op\ with a chiral primary operator $\cO_{\Delta,k}$ and with a Wilson loop $W_{\theta_0,\psi_0}$ we did the 
calculation in the three different realizations of the surface operator \op\ and found 
remarkable agreement. The leading behavior agreed between all three regimes. Particularly 
exciting is the case of the Wilson loop, whose expectation value depends in general on 
the parameters of the surface operator $(N_l,\,\beta_l,\,\gamma_l,\,\alpha_l)$, 
those in the Wilson loop $(\theta_0,\,\psi_0)$ and the 't Hooft coupling $\lambda$.

In all these cases we restricted to a purely classical calculation on the gauge 
theory side, but we expect that those correlators will receive some quantum corrections. 
Remarkably, the computations performed in the probe approximation \probcpofin\ and using the 
``bubbling" 
supergravity description precisely reproduce the semiclassical gauge theory result \correlafin.
Moreover, we have  found that   the correlator of \op\ with some local operators
 have extra terms in \probcpofin \finalbubbles\ 
which are of higher order in the 't Hooft coupling  compared to the semiclassical gauge theory result 
\correlafin, and that nicely organize into a perturbative expansion in $\lambda$. Given these results,  our expectation is that there are perturbative corrections to \correlafin\ which, nevertheless, 
  terminate at   at finite order in the expansion in $\lambda$. In particular, our results suggest that the correlator of \op\ with a chiral primary operator $\cO_{\Delta,k}$ terminate at order  \foot{Note that $\Delta-|k|$ is always  even, so we get a standard perturbation expansion in the 't Hooft coupling $\lambda$.}  $\lambda^{(\Delta-|k|)/2}$,  reproducing 
\probcpofin \finalbubbles. It would be interesting to reproduce this structure of the corrections directly from the path integral description of surface operators \op.

In the case of the Wilson loop $W_{\theta_0,\psi_0}$ we saw the appearance of   terms of the form 
$\pm\sqrt{\lambda N_l/N}$ \metstringmatch, which we attribute to the extra scalar 
$\phi^1$ in \gaugeWL. 
Again, we did not pursue the exact perturbative calculation of its effect, but we 
anticipate that it will give some modification of the matrix model of 
\EricksonAF\ and \DrukkerRR, perhaps along the lines of \SemenoffXP\OkuyamaJC\SemenoffAM.

There exists a definition for the gauge theory in the presence of the surface operator \op\ 
also beyond the classical level. The path integral has to be done in the presence of the 
prescribed singularity. But concrete calculations have never been performed with that 
prescription (prior to this paper no calculations have been done even in the classical 
limit). Our results raise the prospects of being able to do such calculations in practice 
and in the supersymmetric cases considered in this paper possibly improving on our 
results and  computing all  quantum corrections. It would be interesting to use localization
techniques to study whether the path integral of $\cN=4$ SYM in the presence of a surface operator
\op\  becomes that of a reduced model, extending the recent result of Pestun \PestunRZ, who has derived the matrix model in 
\EricksonAF\DrukkerRR\  by analyzing the $\cN=4$ SYM path integral  in the presence of a  supersymmetric circular Wilson loop.

\lref\ucsb{T.~Okuda and D.~Trancanelli, to appear.}
\lref\inprogress{
J.~Gomis, S.~Matsuura, T.~Okuda and D.~Trancanelli, to appear.}

The fact that a supersymmetric surface operator \op\ in $\cN=4$ SYM has multiple 
realizations where calculations can be done with precise agreement has an   
analog in the case of supersymmetric Wilson loops. These can be described
in a variety of ways, perturbatively in $\cN=4$ SYM \EricksonAF\DrukkerRR,
as strings in $AdS_5$  \ReyIK\MaldacenaIM\DrukkerZQ\BerensteinIJ, 
as a configuration of D3-branes \DrukkerKX\GomisSB\GomisIM\ or
as a configuration of D5-branes \YamaguchiTQ\GomisSB, 
and finally as  asymptotically \ads\ ``bubbling'' supergravity backgrounds 
\YamaguchiTE\LuninXR\DHokerFQ. 
This paper aspires to be a first step in generalizing the impressive 
series of comparisons done for Wilson loops between gauge theory calculations and
probe calculations.  Much dynamical information can also be extracted from the ``bubbling" geometries 
for Wilson loops \ucsb\inprogress, extending the results obtained for Wilson loops 
\lref\OoguriBV{
  H.~Ooguri and C.~Vafa,
  ``Knot invariants and topological strings,''
  Nucl.\ Phys.\  B {\bf 577}, 419 (2000)
  [arXiv:hep-th/9912123].
}
 in the context of topological string theory \OoguriBV\ using ``bubbling" Calabi-Yau's
\lref\GomisMV{
  J.~Gomis and T.~Okuda,
  ``Wilson loops, geometric transitions and bubbling Calabi-Yau's,''
  JHEP {\bf 0702}, 083 (2007)
  [arXiv:hep-th/0612190].
}
\lref\GomisKZ{
  J.~Gomis and T.~Okuda,
  ``D-branes as a Bubbling Calabi-Yau,''
  JHEP {\bf 0707}, 005 (2007)
  [arXiv:0704.3080 [hep-th]].
}
\GomisMV\GomisKZ.

\lref\HananyIE{
  A.~Hanany and E.~Witten,
  ``Type IIB superstrings, BPS monopoles, and three-dimensional gauge
  dynamics,''
  Nucl.\ Phys.\  B {\bf 492}, 152 (1997)
  [arXiv:hep-th/9611230].
}
\lref\KarchCT{
  A.~Karch and L.~Randall,
  ``Locally localized gravity,''
  JHEP {\bf 0105}, 008 (2001)
  [arXiv:hep-th/0011156].
}
\lref\GomisCU{
  J.~Gomis and C.~Romelsberger,
  ``Bubbling defect CFT's,''
  JHEP {\bf 0608}, 050 (2006)
  [arXiv:hep-th/0604155].
}
\lref\DHokerXY{
  E.~D'Hoker, J.~Estes and M.~Gutperle,
  ``Exact half-BPS Type IIB interface solutions I: Local solution and
  supersymmetric Janus,''
  JHEP {\bf 0706}, 021 (2007)
  [arXiv:0705.0022 [hep-th]].
}
\lref\DeWolfePQ{
  O.~DeWolfe, D.~Z.~Freedman and H.~Ooguri,
  ``Holography and defect conformal field theories,''
  Phys.\ Rev.\  D {\bf 66}, 025009 (2002)
  [arXiv:hep-th/0111135].
}
\lref\ClarkSB{
  A.~B.~Clark, D.~Z.~Freedman, A.~Karch and M.~Schnabl,
  ``The dual of Janus \break ($(<:) \leftrightarrow (:>)$) an interface CFT,''
  Phys.\ Rev.\  D {\bf 71}, 066003 (2005)
  [arXiv:hep-th/0407073].
}
\lref\KimDJ{
  C.~Kim, E.~Koh and K.~M.~Lee,
  ``Janus and Multifaced Supersymmetric Theories,''
  arXiv:0802.2143 [hep-th].
}
\lref\GaiottoSA{
  D.~Gaiotto and E.~Witten,
  ``Supersymmetric Boundary Conditions in $\cN=4$ Super Yang-Mills Theory,''
  arXiv:0804.2902 [hep-th].
}
\lref\GaiottoSD{
  D.~Gaiotto and E.~Witten,
  ``Janus Configurations, Chern-Simons Couplings, And The Theta-Angle in N=4
  Super Yang-Mills Theory,''
  arXiv:0804.2907 [hep-th].
}
\lref\BerensteinKK{
  D.~Berenstein,
  ``A toy model for the AdS/CFT correspondence,''
  JHEP {\bf 0407}, 018 (2004)
  [arXiv:hep-th/0403110].
}

It would be interesting to perform similar computations to the ones in this paper for the surface operators of order type constructed in \BuchbinderAR. One can also consider maximally supersymmetric domain walls in $\cN=4$ SYM. For these the gauge theory realization is given in terms of the low energy limit of the Hanany-Witten like setup involving D3, D5 and NS5 
branes  \HananyIE, the probe brane description is given by the branes in \ads\ found by Karch and Randall \KarchCT, and the ``bubbling'' description is given by the smooth $Osp(4|4)$ invariant supergravity solutions found in \GomisCU\LuninXR\DHokerXY. The gauge theory description of supersymmetric domain walls in $\cN=4$ has been considered in 
\DeWolfePQ\ClarkSB\KimDJ, and a recent systematic study of supersymmetric boundary conditions has appeared in 
\GaiottoSA\GaiottoSD.

The maximally  supersymmetric operators  in $\cN=4$ SYM are of  great interest. Unlike the 
vacuum, there are many of them, with their  classification   still not complete. Yet, they 
posses a large degree of supersymmetry, sixteen supercharges, and therefore their 
dynamics are under good control, in some cases to all orders in perturbation theory. 

The maximally supersymmetric local operators in $\cN=4$  SYM are captured by a reduction from the four dimensional gauge theory to a matrix quantum mechanics of a single Hermitean matrix
\BerensteinKK, and are classified in terms of Young tableau.  
This matrix model can then be solved in terms of free fermions whose
phase space has been identified  by LLM in terms of boundary conditions for the 
``bubbling'' supergravity description \LinNB. 

The maximally supersymmetric Wilson loops in $\cN=4$ SYM are also of great interest and 
unlike the case of local supersymmetric operators, circular Wilson loops get 
contributions from all orders in perturbation theory.
In fact the circular Wilson loop in $\cN=4$ SYM is captured by a
zero-dimensional matrix model \EricksonAF\DrukkerRR, a result that has been derived 
recently using localization   \PestunRZ. This
matrix model reproduces correctly the expectation value of the loop
as calculated by strings, D-branes \DrukkerKX\YamaguchiTQ\HartnollIS\ and also  reproduces
\lref\OkudaKH{
  T.~Okuda,
  ``A prediction for bubbling geometries,''
  JHEP {\bf 0801}, 003 (2008)  \break
  [arXiv:0708.3393 [hep-th]].
}
properties \OkudaKH\ of the ``bubbling'' supergravity description \ucsb\inprogress.
Modifications of the matrix model 
also encode the interaction of Wilson loops with local operators \OkuyamaJC.

It is tempting therefore to try to see if there exists a matrix model that
captures the essential properties of the maximally supersymmetric surface operators.
We speculate in the following paragraphs over a possible candidate. It would be  interesting to use localization to derive the reduced model capturing the properties of maximally supersymmetric surface operators in $\cN=4$ SYM.

Let us stress from the onset that it is not completely clear what properties
this matrix model should capture. In the case of the supersymmetric circular Wilson loop,  it
calculates its expectation value, which may translate to some conformal
anomaly in our case. Alas we saw that the maximally supersymmetric surface operators \op\ do
not have a conformal anomaly. But one property that we do expect it
to capture is the classification of surface operators in terms of a Levi
group $L=\sum_{l=1}^MU(N_l)\subset U(N)$ and the parameters
$(\alpha_l,\beta_l,\gamma_l,\eta_l)$ with $l=1,\cdots,M$.

\lref\KazakovJI{
V.~A.~Kazakov, I.~K.~Kostov and N.~A.~Nekrasov,
``D-particles, matrix integrals and KP hierarchy,''
  Nucl.\ Phys.\  B {\bf 557}, 413 (1999)
  [arXiv:hep-th/9810035].
}

\lref\HoppeXG{
  J.~Hoppe, V.~Kazakov and I.~K.~Kostov,
``Dimensionally reduced SYM$_4$ as solvable matrix quantum mechanics,''
  Nucl.\ Phys.\  B {\bf 571}, 479 (2000)
  [arXiv:hep-th/9907058].
}

One can consider a zero-dimensional complex matrix model with only
a commutator-squared action
\eqn\mmguess{
Z=\int \cD\Phi\exp\left({1\over g_{YM}^2}\Tr[\Phi,\,\bar\Phi]^2\right)\,.
}
A vacuum of this matrix model is given by a matrix that commute with
its complex conjugate. Such a matrix can then be diagonalized with
eigenvalues $\beta_l+i\gamma_l$, and the remaining symmetry, which
depends on the degeneracy of those eigenvalues, would be exactly such a
Levi group $L=\sum_{l=1}^MU(N_l)$.

This matrix model has an interesting feature.  Due to the lack of a
quadratic term in the action, when the path integral \mmguess\ is written in terms of its eigenvalues including the proper
gauge fixing, the usual Van der Monde determinant is exactly cancelled by a similar
contribution from the Fadeev-Popov determinant \KazakovJI\HoppeXG. 

Normally the
eigenvalues are repelled from each-other due to the Van der Monde determinant,
and this repulsion can then be captured by assigning the eigenvalues
Fermi statistics. For our matrix model, there is no such eigenvalue repulsion and the vacua may
well consist of a large number of coincident eigenvalues.

It is therefore natural to assign to the eigenvalues of the matrix model \mmguess\ Bose statistics, which
nicely captures the structure of boundary conditions of the ``bubbling" solutions for surface operators \op.
The structure of the ``bubbling" supergravity solutions for the surface operators \op\ is
 similar to that of the maximally supersymmetric   local operators in $\cN=4$, where there is also
a 3-dimensional space $X$, with coordinates $(x_1,\,x_2,\,y)$. The main difference between the ``bubbling" solutions for local operators and surface operators is, however,  in the boundary conditions 
for the Laplace equation \laplace, which determines the metric and five-form.

In the case of local operators the supergravity data which determines  
the solution is captured by binary information on the $y=0$ plane,
which is  interpreted \LinNB\ as the occupation number of the free fermions. 
In our
case, corresponding to maximally supersymmetric surface operators \op,  the data 
which determines  the solution is given by the position of point-like sources away
from $y=0$ (see Figure 1). Projecting this picture on the $y=0$ plane, it can be
interpreted as a distribution of {\it bosons}. The distance in the $y$ direction 
which represents the number of coincident D3-branes (or eigenvalues) 
through $y_l^2=N_l/N$, is interpreted as their occupation number and is indeed
quantized in terms of positive integers. 
The ``bubbling" solution for local operators is encoded in terms of fermionic eigenvalues, while the
 ``bubbling" solution for  surface operators  is encoded in terms of bosonic eigenvalues, a property captured by the matrix model \mmguess.

One can enrich the matrix model \mmguess\ by inserting  other  observables in the integrand made out of the fields $\Phi$ 
and $\bar\Phi$, such as chiral primary operators. At least at the semiclassical level this matrix model
captures the correlators of the surface operator \op\ with chiral primary operators.

While this picture is satisfyingly consistent, the matrix model  \mmguess\ has some
shortcomings in describing the surface operators \op. For example, it does not
capture all the details of the vacuum structure. In particular, it does not capture the 
parameters $\alpha_l$ and
$\eta_l$ needed to fully characterize a surface operator \op, and that enter in the correlators od \op\ with the Wilson and 't Hooft operators.
 
  One may be even more ambitious and hope to find a reduced model that 
captures the interactions between the different objects considered in this 
paper, including Wilson and 't Hooft loops. An example to keep in mind is the normal matrix model,  which in 
\OkuyamaJC\ was shown to reproduce the interaction of maximally supersymmetric local operators 
and maximally supersymmetric Wilson loops. 

A possible avenue is to replace the 0-dimensional
matrix model \mmguess\ with matrix quantum mechanics, as is the case
for the local operators. Given that instead of the theory on $S^3\times R$
we are dealing here with $\cN=4$ SYM om $AdS_3\times S^1$,  we propose to 
Wick-rotate and compactify the time direction. We propose to  consider matrix
quantum mechanics of a single complex matrix at finite temperature
\eqn\gmqmlag{
L=\Tr\left[{1\over2}|D_\psi\Phi|^2
-{1\over2}|\Phi|^2+{1\over4}[\Phi,\,\bar\Phi]^2\right].
}
Here the $\psi$ direction is compact and the 
covariant derivative is $D_\psi\Phi=\partial_\psi\Phi-i[A_\psi,\,\Phi]$. 
In a vacuum of this theory,  one can choose a gauge where $A_\psi$ is diagonal, 
and it would then be characterized by the parameters $\alpha_l$ in \holo\ that label a surface operator \op.

In this matrix model it is easy to define, in addition to the local operators, 
a Wilson loop operator wrapping the $\psi$ direction, which would be similar to 
\gaugeWL, at least for $\theta_0=\pi/2$. More degrees of freedom would be required 
in order to capture the dynamics of the extra scalar field $\phi^1$. We leave these 
speculations for future investigation.
\medskip\medskip\medskip\medskip\medskip
\bigbreak\bigskip\noindent{\bf Acknowledgements}\nobreak
\medskip 
We are very grateful to Bartomeu Fiol for his participation in the
initial stages of this project. We would also like to thank Ofer Aharony, 
Evgeny Buchbinder, Alex Buchel,
M{\aa}ns Henningson, Volodia Kazakov, Oleg Lunin, Rob Myers, Takuya Okuda, Vasily Pestun,
Toine Van Proeyen, Riccardo Ricci, Adam Schwimmer,
Stefan Theisen and Diego Trancanelli for useful discussions and
Kostas Skenderis and Marika Taylor for correspondence.

N.D and J.G would like to thank the Galileo Galilei Institute, Florence,
where this collaboration was formed, and a year later completed,
for its hospitality and the infn for partial financial support.
N.D  would also like to acknowledge the warm hospitality of
The Perimeter Institute and Weizmann Institute during the course of
this work.
S.M  would like to acknowledge the warm hospitality of
MCTP at the University of Michigan.

Research at Perimeter Institute is supported by the Government
of Canada through Industry Canada and by the Province of Ontario through
the Ministry of Research and Innovation. J.G.  also acknowledges further  support by an NSERC Discovery Grant.

\vfill\eject

\appendix{A}{Spherical Harmonics}

In this Appendix we summarize some of the properties of $SO(6)$  spherical harmonics (see \SkenderisYB\ for more details).
The scalar and   vector  $SO(6)$  spherical harmonics satisfy the following equations,
\eqn\spherical{\eqalign{
  \triangle Y^{I_1}&=\Lambda^{I_1}Y^{I_1},~~~~~
  \Lambda^{I_1}=-\Delta(\Delta+4),~~~~~~~~~~\Delta=0,1,2,\cdots\cr
\triangle Y^{I_5}_{a}&=\Lambda^{I_5}Y^{I_5}_{a},~~~~~\Lambda^{I_5}=-(\Delta^2+4\Delta-2),~~~~\Delta=1,2,\cdots\,,
}}
where $\triangle$ is Laplacian on $S^5$. They are normalized as follows
\eqn\normalization{\eqalign{
\int Y^{I_1}Y^{J_1}&=\pi^3 z(\Delta)\delta^{I_1J_1},\cr
\int Y^{I_5}_{a}Y^{J_5a}&=\pi^3 z(\Delta)\delta^{I_5J_5},
}}
where 
\eqn\zz{
z(\Delta)={1\over 2^{\Delta-1}(\Delta+1)(\Delta+2)}.
}
We consider the following metric \metricsphere\ on $S^5$
\eqn\spherecoord{
ds^2=\cos^2{\theta}d\Omega_3+d\theta^2+\sin^2{\theta}d\phi^2,
}
where $\theta\in [0,\pi/2]$ and $\phi\in  [0,2\pi]$.

The $SO(4)$ invariant scalar spherical harmonics of $SO(6)$ are
given by
\eqn\scalarsphe{
Y^{\Delta,k}(\theta,\phi)=c_{\Delta,k}\,y^{\Delta,k}(\theta)e^{ik\phi},
}
where $\Delta$ is a non-negative integer,
$k\in (-\Delta, -\Delta+2, \cdots,\Delta-2,\Delta$),
$c_{\Delta,k}$ is a normalization constant and $y^{\Delta,k}$ is given by
a hypergeometric function 
\eqn\smally{
y^{\Delta,k}=\sin^{|k|}{\theta}\
{}_2F_1\left(-{1\over2}(\Delta-|k|),2+{1\over2}(\Delta+|k|),1+|k|;\sin^{2}{\theta}  \right).
}

For the computations in this paper, we need the following scalar spherical harmonics
\eqn\scasphelist{\eqalign{
Y^{2,0}&={1\over2\sqrt{3}}(3\sin^2{\theta}-1),\cr
Y^{\Delta,\pm \Delta}&={1\over2^{\Delta/2}}\sin^{\Delta}{\theta}e^{\pm i\Delta\phi},\cr
Y^{3,\pm 1}&={\sqrt{3}\over4}\sin{\theta}(2\sin^2{\theta}-1)e^{\pm i\phi}
}}
and   vector spherical harmonic 
\eqn\scasphelist{\eqalign{
Y^1&={1\over \sqrt{2}}\sin^2{\theta}d\phi.}}

In order to write down the explicit form of the dimension $\Delta=2,3$ $SO(4)$ invariant chiral primary operators  \opers\ we write the spherical harmonics in \scasphelist\ using the following embedding coordinates 
\eqn\embcoord{\eqalign{
X^i&=\cos\theta\, \Theta_i ~~~~~ i=1,\cdots,4\cr
X^5&=\sin\theta \cos\phi,\cr
X^6&=\sin\theta \sin\phi,
}}
where $\sum_{i=1}^{4}\Theta^2_i=1$.

Evaluating the harmonics at $\theta=\pi/2$ we get that $Y^{\Delta,k}(\theta=\pi/2,\phi)=C_{\Delta,k}e^{ik\phi}$,
where 
\eqn\normaprico{
C_{2,0}=1/\sqrt{3},\qquad  C_{2,\pm2}=1/2, \qquad C_{3,\pm1}=\sqrt{3}/4 , \qquad
C_{3,\pm3}=1/2\sqrt2}
are obtained from the normalization condition \normalization.

\appendix{B}{Supersymmetry of Wilson Loops}

We review here the supersymmetry calculation for the
Wilson loop \gaugeWL, as done in \DrukkerGA
  and show that they are compatible with those preserved by the surface operator \op. We follow the notations in the Appendix of \GomisFI,
using 10-dimensional gamma matrices
such that the variation of the two components of the gauge
field gives terms with the gamma matrices $\gamma^2$ and
$\gamma^3$ and of the two scalars in $\Phi$ and $\bar\Phi$ gives
$\gamma^8\pm i\gamma^9$. In addition the variation of the
extra scalar $\phi^1$ gives $\gamma^4$.

The supersymmetry variation of the Wilson loop $W_{\theta_0,\psi_0}$ \gaugeWL\ gives an expression proportional to
\eqn\susywilson{\eqalign{
&\Big(\sin\psi(-i\gamma^2+\sin\theta_0\sin\psi_0\gamma^8
-\sin\theta_0\cos\psi_0\gamma^9)
\cr&
+(\cos\psi(i\gamma^3+\sin\theta_0\cos\psi_0\gamma^8
+\sin\theta_0\sin\psi_0\gamma^9)
+\cos\theta_0\gamma^4\Big)
\big(\epsilon_1+\cos\psi\gamma^2\epsilon_2
+\sin\psi\gamma^3\epsilon_2\big).
}}
Here $\epsilon_1$ are the parameters of the Poincar\'e supersymmetry 
transformations and $\epsilon_2$ are the parameters of the  of the conformal supersymmetry transformations.
The supersymmetries preserved by the loop are those where the
above expression \susywilson\ vanishes for all $\psi$. One can therefore expand  \susywilson\  in
Fourier modes and get different equations for each Fourier component: 1, $\sin\psi$,
$\cos\psi$, $\sin 2\psi$ and $\cos 2\psi$.

It turns out that requiring all these projectors to annihilate
the spinors is a consistent set of equations which have 8 independent
solution (for $\theta_0=0$ there are 16). Therefore such a Wilson
loop is quarter BPS.

\if 
It turns out that requiring all these Fourier components  to annihilate
the supersymmetry parameters   $\epsilon_1$ and $\epsilon_2$ is a consistent set of equations which have four independent
solutions both for $\epsilon_1$ and $\epsilon_2$  (the case when  $\theta_0=0$ there are eight nonvanishing components for $\epsilon_1$ and $\epsilon_2$ ). Therefore,  the  Wilson
loop $W_{\theta_0,\psi_0}$  preserves one quarter of the thirty-two supersymmetries.
\fi

These equations in general relate $\epsilon_1$ and $\epsilon_2$,
but we can separate them and find the following equation that
is satisfied by both spinors
\eqn\susyWL{
\sin\theta_0\big[\sin\psi_0\,(\gamma^{28}+\gamma^{39})
+\cos\psi_0\,(\gamma^{38}-\gamma^{20})\big]\epsilon=0\,.
}

The supersymmetries unbroken by the surface operator in the
plane transverse to the loop are the solutions to the equations
\GomisFI\ (and see also the next appendix)
\eqn\surfacesusys{
\gamma_{2389}\,\epsilon=-\epsilon\,,
}
with either $\epsilon_1$ or $\epsilon_2$.
It is easy to check that any solution to \surfacesusys\
automatically solves \susyWL. Therefore the presence of the
surface operator \op\ does not break any of the supersymmetries 
left unbroken by the Wilson loop \gaugeWL, they share eight supercharges.

In the special case of the maximally supersymmetric loop, for $\theta_0=0$,
the surface operator does break half of the supersymmetries of the Wilson loop,
so the combined system is quarter supersymmetric, {\it i.e.} preserving
eight supercharges, like the generic case.

\appendix{C}{Supersymmetry of  Probe D3-Brane}

In this Appendix, we study the supersymmetries left unbroken by a
probe D3-brane corresponding  to
a surface operator \op. To avoid the complications of Euclidean
supersymmetry, we consider \op\ supported on $\Sigma=R^{1,1}$ in $R^{1,3}$.
We take the metric on  $AdS_5\times S^5$  to be
\eqn\psmetric{\eqalign{
ds^2=&{d\omega^2 \over \omega^2}
+\omega^2(-dx_0^2+dx_1^2+dx_2^2+dx_3^2)
+d\theta_0 ^2+\sum_{j=1} ^{3} \prod _{i=0} ^{j-1}\sin^2{\theta_i}\, d\theta _j ^2+
\prod _{i=0} ^{4}\sin^2{\theta_i}\, d\phi ^2.\cr
}}
The Killing spinor of $AdS_5\times S^5$ in this coordinate system is given by \SkenderisVF
\eqn\killing{
\epsilon=
\left[-\omega^{-{1\over2}}\gamma_4\,h(\theta_i,\phi)+
\omega^{{1\over2}}h(\theta_i,\phi)x_m\gamma^m)\right]\eta_2
+\omega^{{1\over2}}h(\phi_i,\phi)\eta_1,
}
where $\gamma_\mu$ are the tangent space gamma matrices and
\eqn\eichi{
h(\phi_i,\phi)=
e^{{1\over2}\theta_0\gamma_{45}}
e^{{1\over2}\theta_1\gamma_{56}}
e^{{1\over2}\theta_2\gamma_{67}}
e^{{1\over2}\theta_3\gamma_{78}}
e^{{1\over2}\phi\gamma_{89}}.
}
$\eta_1$ and $\eta_2$ are constant ten dimensional complex spinors
that have positive and negative ten dimensional chirality respectively,
\eqn\chiral{
\gamma_{11}\eta_1=-\eta_1,~~~~~~\gamma_{11}\eta_{2}=\eta_2,
}
which also satisfy
\eqn\epseps{
\eta_1=(1-i\gamma_{0123})\epsilon _1,~~~
\eta_2=(1+i\gamma_{0123})\epsilon _2,
}
where $\epsilon_1$ and $\epsilon_2$ are ten dimensional Majorana-Weyl fermions with positive and negative chirality.
At the boundary, $\epsilon_1$ is identified with a Poincar\'e supersymmetry parameter and
$\epsilon_2$ with a conformal supersymmetry    parameter of the dual gauge theory.

The embedding of the supersymmetric  probe D3 brane \yrsolution\ is given
in this coordinate system by  
\eqn\emb{\eqalign{
r={\sinh u_0\over\omega}\,,\qquad
\theta_{i}={\pi\over 2}\,,\qquad
\psi+\phi=\phi_0\,,
}}
where $x_2+i x_3=r\,e^{i\psi}$.
In this metric and embedding, the induced metric on the D3-brane
is $AdS_3\times S^1$
\eqn\mmetric{
ds^2
=\cosh^2u_0\left({d\omega^2\over \omega^2}
+{\omega^2\over\cosh^2u_0}(-dx_0^2+dx_1^2)
+d\phi^2\right).
}
\smallskip
The supersymmetries that are preserved by the D3-brane
are those that satisfy
\eqn\kapproj{
\Gamma_{\kappa}\epsilon |_{D3}=\epsilon |_{D3},
}
where $\epsilon$ is the Killing spinor of the $AdS_5\times S^5$
background given in \killing, and
$\Gamma_{\kappa}$ is the $\kappa$-symmetry projection matrix
\eqn\kapa{
\Gamma_{\kappa} =-i\cL^{-1}_{DBI}\,
\partial_0x^\mu\,\partial_1 x^\nu\,\partial_\omega x^\rho\,
\partial_\phi x^\sigma\,\Gamma_{\mu\nu\rho\sigma}\,.}
 For our D3 embedding \emb
\eqn\gammafour{\eqalign{
i\cL_{DBI}\Gamma_\kappa&=
\Gamma_{01\omega\phi}+{\sinh^2u_0\over\omega^3}\,\Gamma_{0123}
-{\sinh u_0\over\omega^2}\sin(\phi-\phi_0)
(\omega\Gamma_{01\omega2}-\Gamma_{013\phi})\cr
&\qquad
{}-{\sinh u_0\over\omega^2}\cos(\phi-\phi_0)
(\omega\Gamma_{01\omega3}+\Gamma_{012\phi}).
}}
The gamma matrices appearing in the last two equations are related
to the  tangent space $\gamma$ matrices by
\eqn\tamgamma{\eqalign{
\Gamma_{\mu}=\omega\,\gamma _{\mu}\,,\qquad
\Gamma_{\omega}={1 \over \omega}\,\gamma _4\,,\qquad
\Gamma_{\phi}=\gamma_{9}\,.
}}
Using $\cL_{DBI}=\omega\cosh^2u_0$ we may write $\Gamma_{\kappa}$ as
\eqn\gammmma{\eqalign{
\Gamma_{\kappa}&=-{i \over \cosh^2u_0}\Big(
(\gamma_{0149}+\sinh^2u_0\,\gamma_{0123})
-\sinh u_0\sin(\phi-\phi_0) (\gamma_{0142}-\gamma_{0139})\cr
&\hskip1in
-{\sinh u_0}\cos(\phi-\phi_0)(\gamma_{0143}+\gamma_{0129})\Big).
}}
After some  algebra we get from \kapproj
\eqn\superkilling{\eqalign{
(1+\gamma_{2389})\epsilon_1&=0,\cr
(1+\gamma_{2389})\epsilon_2&=0.}}
The preserved supersymmetries \superkilling\
 are the same as those preserved by a maximally supersymmetric  surface operator $\ope$ in $\cN=4$ SYM (see for e.g. \GomisFI).

\appendix{D}{Gauge Transformation on $V$}

As mentioned in the main text,
the one-form \oneform\ is determined up to an exact form by solving the equation $dV={1\over y} *_{X}dz$.
In order to get a manifestly  asymptotically $AdS_5\times S^5$ geometry,
we need to add to V \oneform\ an exact one-form 
\eqn\metriccorrec{
V\rightarrow V+d\omega.
}

The exact form is given up to ${\cal O}({1/ R^4})$
\eqn\exofomeg{\eqalign{
\omega &=-{M-1\over2}\alpha+\sum_{l=1}^M \left(
{x_{l,2}\cos{\phi}-x_{l,1}\sin{\phi}\over 2\sin{\theta}R}
\right)\hskip-3pt-\hskip-5pt \sum_{l=1}^M \hskip-3pt\left({\left(x^2 _{l,1}-x^2 _{l,2}\right)\sin{2\alpha}
-2x_{l,1}x_{l,2}\cos{2\alpha}\over 4\sin^2\theta R^2}\hskip-2pt\right)\cr
&-\sum_{l=1}^M
\left({x_{l,1} ^3\sin{3\alpha}+x_{l,2} ^3\cos{3\alpha}
-3x_{l,1} ^2x_{l,2}\cos{3\alpha}-3x_{l,1}x_{l,2} ^2\sin{3\alpha}
\over6\sin^3\theta R^3}\right)+{\cal O}\left({1\over R^4}\right).
}}
where we have used the coordinates in \rela.
In computing the correlation functions of surface operators \op\ with local operators using the ``bubbling" supergravity solution, we have used this expression for $V$.

\listrefs

\end